\documentclass[]{aa}
\usepackage{graphicx,natbib}
\newcommand{\lya}{Ly$\alpha$}
\newcommand{\ha}{H$\alpha$}
\newcommand{\ecs}{erg cm$^{-2}$ s$^{-1}$}
\newcommand{\kms}{km s$^{-1}$}
\newcommand{\nbo}{NB\,0.38}
\newcommand{\nbir}{NB\,2.07}
\newcommand{\fone}{$\mathcal{F}$1}
\newcommand{\ftwo}{$\mathcal{F}$2}
\newcommand{\rg}{PKS~1138$-$262}
\newcommand{\rgs}{1138$-$262}
\newcommand{\sfrha}{SFR$_{\rm H\alpha}$}
\newcommand{\sfruv}{SFR$_{\rm uv}$}
\newcommand{\sfrlya}{SFR$_{\rm Ly\alpha}$}
\newcommand{\my}{M$_\odot\,$yr$^{-1}$}
\begin{document}
  \title{A search for clusters at high redshift}
  \subtitle{III. Candidate \ha\ emitters and EROs in the
  PKS~1138$-$262 proto-cluster at $z=2.16$}

  \author{J.~D. Kurk \inst{1} \and L. Pentericci \inst{2} \and
           H.~J.~A. R\"ottgering \inst{1} \and G.~K. Miley \inst{1}}

  \institute{Sterrewacht Leiden, P.O. Box 9513, 2300 RA, Leiden, 
             The Netherlands               
        \and
             Max-Planck-Institut f\"ur Astronomie, K\"onigstuhl 17,
             D-69117, Heidelberg, Germany}

%   \offprints{J.D. Kurk, \email{kurk@strw.leidenuniv.nl}}
  \offprints{J.D. Kurk (kurk@strw.leidenuniv.nl)}

  \date{Received date / Accepted date}

\abstract{In this paper we present deep VLT multi wavelength imaging
observations of the field around the radio galaxy
\object{PKS~1138$-$262} aimed at detecting and studying a (potential)
proto-cluster centered at this radio source.  \rg\ is a massive galaxy
at $z = 2.16$, located in a dense environment as indicated by optical,
X-ray and radio observations.  We had already found an over-density of
\lya\ emitting galaxies in this field, consistent with a proto-cluster
structure associated with the radio galaxy.  In addition, we find 40
candidate \ha\ emitters that have nominal rest frame equivalent width
$> 25$ \AA\ within 1.8 Mpc and 2000 \kms\ of the radio galaxy.
Furthermore, we find 44 objects with $I - K > 4.3$.  This number of
extremely red objects (EROs) is about twice the number found in blank
field ERO surveys, suggesting that some EROs in this field are part of
the proto-cluster.  The density of \ha\ emitters and extremely red
objects increases towards the radio galaxy, indicating a physical
association.  From comparisons with other $K$ band, ERO, \ha\ and
\lya\ surveys, we conclude that \rg\ is located in a density peak
which will evolve into a cluster of galaxies.   }

\maketitle
%
%________________________________________________________________

\section{Introduction}
The search for clusters at high redshift has two main incentives:
distant clusters can be used to constrain cosmological models and they
provide a reservoir of high redshift galaxies, which can be used to
study galaxy formation and evolution.

According to hierarchical clustering theories, clusters form by the
gravitational amplification of primordial density fluctuations. In a
low density universe, fluctuations cease to grow after a redshift $z
\sim (\Omega_0^{-1} -1)$ \citep{pee80}, resulting in a cluster
population that evolves very slowly at low redshift. In an $\Omega_0 =
1$ universe density fluctuations continue to grow even at the present
epoch, implying that the cluster population would still be evolving
rapidly \citep{eke96}. The detection of even a single distant massive
cluster, such as MS 1054$-$03 at $z = 0.83$, constrains the parameters
($\Omega_0, \sigma_8$) of cosmological models \citep{bah98, don98}.

The study of galaxies in nearby and distant clusters provides
strong constraints on their evolution and formation. It has been
shown that clusters at high redshift can contain a different galaxy
population mix than those at low redshift.  \citet{but84} found that
at $0.1 < z < 0.5$, compact clusters have significant numbers of blue
galaxies, the fraction increasing with redshift, while at $z < 0.1$
cluster cores are essentially devoid of these. This effect can be
explained if we assume a process in which spirals lose their gas and
ability to form stars or are converted into early type galaxies in the
course of their evolution. Massive ellipticals, however, dominating
cluster cores up to $z \sim 1$ seem to form a homogeneous
population.  Even at $z = 1.27$, \citet{dok01a} observe a scatter
of the colour-magnitude relation of cluster galaxies similar to that
in lower redshift clusters.

The tight colour-magnitude relation observed out to high redshifts can
be explained by several galaxy formation scenarios.  \citet{pro99}
propose a passive luminosity evolution model, where galaxies form all
their stars in a single burst at $z = 2$ after a monolithic collapse
\citep[e.g.][]{egg62}.  \citet{dok01b} show that the observations are
also consistent with a scenario in which early-type galaxies are
continuously transformed from spiral galaxies, causing the progenitors
of the youngest, low redshift early-type galaxies to drop out of the
sample at higher redshifts (\emph{progenitor bias}). Models by
\citet{dok01b} show that about half of the early-type galaxies were
morphologically transformed at $z < 1$ and their progenitors may have
had roughly constant star formation rates before transformation.
These models indicate a mean luminosity weighted formation of stars in
early-type galaxies of $z = 2.0$ (for $\Omega_{\rm M}$ = 0.3 and
$\Omega_\Lambda$ = 0.7), consistent with the currently favoured
hierarchical galaxy formation models predicting the merging of smaller
galaxies at high redshift \citep{kau96}.  A third explanation is given
by metallicity differences in bright and faint ellipticals.  A model
by \citet{kau98} which includes hierarchical formation of ellipticals
out of disc galaxies which have formed stars at modest rates and
allows for the ejection of metals out of discs by supernova explosions
predicts the establishment of a mass-metallicity relation among both
late- and early-type galaxies.  In this model, large ellipticals are
more metal-rich because they are formed from the mergers of larger
discs.

The strongest observational constraints on these models come from the
highest redshift data.  Both for the study of cluster and galaxy
evolution a sample of clusters at high redshift is therefore desired.

\medskip

\noindent In recent years, much effort has been invested in the search
for distant clusters, using both optical and X-ray
observations. At $z > 0.5$, it becomes difficult to identify the
projected two-dimensional over-density produced by cluster galaxies,
because large numbers of foreground and background galaxies reduce the
density contrast in the optical wavelength regime.  However, the $J-K$
colour of nearly all galaxies out to $z = 2$ is a simple function of
redshift, because their near infrared light is dominated by evolved
giant stars. The highest redshift cluster found to date \citep[CIG
J0848+4453 at $z = 1.27$]{sta97} has been discovered in a near
infrared field survey as a high density region of objects with very
red $J-K$ colours. Optical spectroscopy confirmed the redshifts of
eight members and a 4.5$\sigma$ ROSAT X-ray detection confirms the
cluster's existence.  Although the detectability of hot cluster gas in
X-rays is severely reduced by cosmological surface brightness dimming,
\citet{ros98} has found within the ROSAT Deep Cluster Survey a cluster
at $z = 1.11$ \citep[RDCS J0910+5422]{sta02} with 9 spectroscopically
confirmed cluster members and a cluster at $z = 1.26$ \citep[RX
J0848.9+4452]{ros99} with 6 cluster members confirmed. Most of the
confirmed galaxies have red colours, consistent with passively evolved
ellipticals formed at high redshift ($z \sim 5$).  The latter cluster
is very close to CIG J0848+4453, with which it might form a
superstructure and is possibly in the process of merging.  The
recently started XMM Large Scale Structure Survey \citep{ref02},
covering over 64 square degrees of sky, should be able to detect
clusters with X-ray luminosities of $2 \times 10^{44}$ erg s$^{-1}$
out to $z = 2$.  Despite the success of the near infrared and X-ray
techniques, it is difficult to push these methods to find $z \gg 1$
over-densities.

\medskip

\noindent A practical way to find clusters and groups of galaxies at
high redshift is to study fields containing luminous radio galaxies.
These can be observed up to the epoch of galaxy formation and
efficiently selected by their steep spectrum in the radio regime
\citep{rot94,lac94,bre00}. The most distant radio galaxy found to date has a
redshift of 5.2 \citep{bre99b}. The host galaxies of powerful radio
sources are amongst the most massive at any redshift
\citep{jar01,bre02} and are associated with $\sim 10^9$ M$_\odot$ BHs
\citep{lac01,mcl02}.

There has long been evidence that powerful radio galaxies at high
redshift (HzRGs, $z > 2$) are located in the center of (forming)
clusters of galaxies.  \citet{yat89} and \citet{hil91} find that the
average environment of 70 powerful classical double radio sources at
$0.15 < z < 0.82$ is that of an Abell 0 cluster, with some in
environments as rich as Abell class 1.  At $z \sim 1$ and higher,
there is also evidence for galaxy over-densities associated with radio
galaxies. \citet{bes00} presents an analysis of the environments of 28
3CR radio galaxies at $0.6 < z < 1.8$. The density of $K$-band
galaxies in these field, their angular cross correlation amplitude and
near infrared colours correspond to the properties of low redshift
Abell richness class 0 to 1 clusters. The author concludes that many,
but not all, powerful radio galaxies at $z \sim 1$ lie in 
cluster environments.  Furthermore, \citet{nak01} have applied a
photometric redshift technique based on five optical and near infrared
images of the field of 3C324 at $z = 1.2$ and identified 35 objects as
plausible cluster members.  The evidence extends even to $z = 3.8$,
where observations in the field of radio galaxy 4C41.17 by
\citet{ivi00} tentatively reveal a number of luminous submm galaxies
over-dense by an order of magnitude as compared to typical fields.

There is also evidence that the environments of HzRGs are dense in
terms of ambient gas, as expected for the centers of (forming)
clusters.  Radio continuum observations of $\sim 70$ radio galaxies at
$z \sim 2$ \citep{car97,pen00b} show that 20 -- 30\% have large ($\geq
1000$ rad m$^{-2}$) radio rotation measures (RMs). These RMs are most
probably due to magnetized, ionized gas local to the radio sources and
are comparable to the RMs of lower redshift radio galaxies which lie
in the centers of dense, X-ray emitting cluster atmospheres
\citep{tay94}.

%\citet{fab01} have observed X-ray emission extending
%well beyond the powerful radio galaxy 3C294 at $z = 1.786$, but the
%analysis of a more recent 200\,ks spectrum favours a mostly
%non-thermal origin of the emission \citep{fab03}.

%One X-ray atmosphere has been observed with \emph{Chandra} around the
%powerful radio galaxy 3C294 at $z = 1.786$ \citep{fab01}. The
%hourglass shaped X-ray emission extends well beyond the radio source
%and its spectrum is fit by a cooling flow model with cooling rates of
%$\sim 400 - 700$ M$_\odot$ yr$^{-1}$.

%Probably, the powerful radio galaxies at all redshifts can be found in
%rich environments which provide enough material to feed the active
%nucleus and have jets which are strong enough not to be hampered by a
%dense environment, as seems to be the case for the local example of
%Cygnus A, which is the brightest member of a cluster of richness class
%$\geq 1$ \citep{owe97}.

\medskip

\noindent Although convincing evidence for high density environments
associated with radio galaxies at $z > 1$ has been demonstrated, the
redshifts of possible cluster members have not been confirmed by
spectroscopy and it is therefore impossible to provide a velocity
dispersion for these structures.  In a program which is currently
being carried out with the VLT, we are targeting the fields of
luminous radio galaxies at $z > 2$ and observe these with the aim of
detecting line emitting galaxies in the associated cluster.  At $z >
2$ the \lya\ line is redshifted into the optical wavelength region,
where we can use narrow band filters ($\sim$ 1\%) to isolate its flux
from the sky background. We have selected ten luminous radio galaxies
at $2.2 < z < 5.2$, for which we will carry out both imaging and multi
object spectroscopy. The survey is progressing very well and has
already produced the discovery of the most distant structure of
galaxies known \citep[at $z = 4.1$,][]{ven02}.
%When finished, we should be able to estimate what fraction of HzRGs is
%located in galaxy over-densities and how the proto-cluster properties,
%such as velocity dispersion and size, and member properties, such as
%masses and star formation rates evolve with redshift. Eventually, we
%will investigate whether the evolution of HzRGs and associated
%over-density peaks are consistent with models of structure formation
%and put constraints on cosmological parameters.

We consider the \lya\ imaging and spectroscopy as a first step to the
characterization of the cluster properties. The \lya\ emitters
in a cluster form only a fraction of the galaxies present and
might not be representative for some of the cluster or galaxy
properties. The program will therefore be followed up by e.g.\ broad
band imaging in several colours. In this paper, we present new optical
and near infrared observations of the field of \rg\ with the
aim of uncovering populations of \ha\ emitting galaxies and EROs in
the structure.

The radio galaxy \rg\ at a redshift of 2.16, was selected from a
compendium of more than 150 $z > 2$ radio galaxies as the optimum
object for beginning a high redshift cluster search. It combines most
of the above mentioned cluster indications with a redshift suitable
for both \lya\ and \ha\ imaging. The magnitude of \rgs\ is the
brightest of all known radio galaxies close to $z = 2$. After
correction for possible non stellar components, the $K$ band magnitude
is 16.8, from which a stellar mass of $10^{12}$ M$_\odot$ was inferred
\citep{pen97}. The radio galaxy possesses a giant ($\sim$ 120 kpc) and
luminous \lya\ nebula, with a wealth of structure: a bright region
associated with the radio jet and filaments extending over $>$ 40 kpc
\citep{pen97,kur00b}. The optical counterpart of the radio galaxy is
extremely clumpy and resolved into many components by the HST
\citep{pen98}. These clumps have properties similar to LBGs. The
morphology of the system is consistent with hierarchical models of
galaxy formation in which the LBG building blocks will merge into a
single massive system, such as the massive galaxies observed at the
centers of some rich clusters. The extremely distorted radio
morphology \citep{car97} is strong evidence that the jets have been
deflected from their original direction by a dense and clumpy
medium. The observed rotation measures of the radio emission (6200 rad
m$^{-1}$, the largest in a sample of 70 HzRGs, see
\citealt{car97,pen00b}) and its steep gradient over the radio galaxy
components also testify that the radio source is embedded in a dense
magnetized medium. Additional evidence for a dense surrounding medium
comes from \emph{Chandra} X-ray observations, which reveal thermal
emission from shocked gas \citep{car02}. The pressure of this hot gas
is adequate to confine the radio source.

Narrow band imaging of redshifted \lya\ emission of a
7\arcmin$\times$7\arcmin\ region around the radio galaxy
\citep[][Paper I]{kur00} and subsequent \lya\ spectroscopy
\citep[][Paper II]{pen00a} revealed 14 \lya\ emitting galaxies and one
QSO. The galaxies have redshifts in the range 2.16$\pm$0.02 with a
velocity dispersion substantially smaller than expected for a random
sample of galaxies selected by the narrow band filter. In addition,
the \emph{Chandra} X-ray observations of the field of \rgs\ have
revealed at least five AGN at $z \sim 2.16$ \citep{pen02}.  On the
basis of the evidence from the radio galaxy properties, the \lya\ halo
and the galaxy over-density, we concluded that the structure of
galaxies surrounding \rg\ is (the progenitor of) a cluster.

\medskip

\noindent The new observations of \rgs\ are reported in
Sect.~\ref{obs}. Detection and photometry of objects in the field of
\rgs\ are presented in Sect.~\ref{cat}. Subsequently, the selection
from these objects of $K$ band galaxies, EROs, candidate \ha\ emitters
and candidate \lya\ emitters is presented in Sect.~\ref{can}. The
properties of the EROs and candidates are analyzed in
Sect.~\ref{prop}.  A discussion of the implications of these results
for the nature of the structure can be found in Sect.~\ref{clus},
which is followed by a summary of the results and conclusions in
Sect.~\ref{sum}.  Throughout this article, we adopt a Hubble constant
of $H_0 = 65$ km s$^{-1}$Mpc$^{-1}$ and a $\Lambda$ dominated
cosmology: $\Omega_{\rm M}$ = 0.3 and $\Omega_\Lambda$ = 0.7.  The
over-densities of galaxies at high redshift, which have not yet
reached virialization and/or a colour-magnitude relation with a red
sequence, but will later form clusters, will be called
\emph{proto-clusters} here.

%__________________________________________________________________

\section{Observations and data reduction}\label{obs}

\subsection{Optical observations}
The observations of \rg\ were carried out with the VLT\footnote{Based
on observations carried out at the European Sou\-thern Observatory,
Paranal, Chile, programmes P63.O-0477(A\&D), P65.O-0324(B), and
P66.A-0597(B\&D).}. With the aim of detecting \lya\ emitting galaxies
at $z = 2.16$, we have observed the field of \rgs\ for half an hour in
$B$ band and four hours in a 2\% narrow band, using FORS1 at Antu
(UT1).  Subsequent multi object spectroscopy of candidate emitters was
also carried out with FORS1, employing three masks with integration
times of 4 to 6 hours.  The optical imaging and spectroscopy
observations are described in detail in Paper I and II.  For an
overview of both old and new observations, see Table \ref{obs_table}.

We have complemented the original optical imaging with broad band
observations in $R$ and $I$, using FORS2 at Kueyen (UT2) in 2001.  The
detector of FORS2 was a Tektronix thinned and anti-reflection coated
CCD with 2048$\times$2048 pixels and a scale of 0\farcs2 per pixel in
standard resolution mode, yielding a field size of $\sim$
6\farcm8$\times$6\farcm8. Six exposures of 5 minutes during
non-photometric conditions were taken through the $R$\_Special filter,
which has a central wavelength of 6550 \AA\ and FWHM of 1650 \AA. The
$R$\_Special filter has higher transmission than the standard Bessel
$R$ filter and its transmission curve is almost symmetrical around the
central wavelength, while the Bessel filter has its peak at 6000 \AA\
and declines towards the red. Six non-photometric exposures of 7.5
minutes were taken in service mode through the Bessel $I$ filter,
which has a central wavelength of 7680 \AA\ and FWHM of 1380
\AA. During visitor time, three weeks later, an additional eighteen
photometric exposures of 5 minutes were observed.  The observations
were made employing a jittering pattern with offsets $<$ 20\arcsec\
between exposures to minimize flat fielding problems and to facilitate
cosmic ray removal. The seeing on the resultant images and the
3$\sigma$ limiting magnitude in a 1\arcsec\ aperture as measured on
the central square arcminute of the combined images is listed in Table
\ref{reduced}.

Image reduction was carried out using the IRAF\footnote{IRAF is
distributed by the National Optical Astronomy Observatories, which are
operated by the Association of Uni\-versities for Research in
Astronomy, Inc., under cooperative agreement with the National Science
Foundation.} reduction package. The individual frames were bias
subtracted, flat fielded with twilight flats and cosmic rays were
removed. The frames were combined using the DIMSUM\footnote{DIMSUM is
a set of scripts to reduce dithered images contributed to IRAF and
written by P.~Eisenhardt, M.~Dickinson, A.~Stanford, J.~Ward and F.\
Valdes.} package. DIMSUM builds a cumulative sky frame from 6 to 10
subsequent unregistered images. Objects in the unregistered frames
were detected with SExtractor \citep{ber96} and masked during the
process of background determination. The obtained sky frames were
subtracted from the images. The image offsets were determined by
measuring the positions of a number ($\sim 20$) of stars on each
frame. Pixels on the CCD which were significantly discrepant in each
sky frame were marked as bad pixels and the exposure time for each
pixel was computed by DIMSUM.  This exposure map is later used as a
weight map for object detection and photometry.  In the last step of
the process, all individual broad band frames were combined by
averaging while identified cosmic rays and bad pixels were omitted.
The registered narrow band images were combined by computing the
average of each pixel stack and rejecting pixels whose intensity
levels were 10$\sigma$ above or below the noise level expected from
the CCD gain and readout noise specifications.

For the flux calibration of the photometric data the standard stars
GD108 \citep{oke90} and LTT4816 \citep{ham92, ham94} were used. The $I$
band data obtained during non-photometric conditions was scaled to the
photometric $I$ band data and the $R$ band data was calibrated using an
older $R$ band image of \rgs\ from \citet{pen97}. Astrometric
calibration was carried out by identifying 18 stars in the USNO-A2.0
catalogue \citep{mon98}, which is tied to the Tycho catalogue
\citep{hog97}. The absolute astrometric accuracy obtained in this way
is $\sim$ 0\farcs2.

Note that the narrow and broad band images obtained in 1999 (Paper I)
were reduced again. This time we used DIMSUM and obtained a
homogeneously reduced set of images in all observed optical and
infrared bands. To overcome differences in geometrical
distortion, all images were mapped to match the $R$ band image.  In
this way, we have obtained a first set of images with their original
spatial resolution.  A second set was made in which the $I$, $B$ and
narrow band ([\ion{O}{ii}]/8000, called \nbo\ from now on) images were
convolved with the kernel required to match their point spread
functions to the $R$ band image, which had the worst seeing conditions
during observations.  The pixel-to-pixel alignment in the final
images is accurate to within a pixel (0\farcs2) over the entire image.

%For multi band photometry (Sect.~\ref{cat}), objects on all images
%have to be astrometrically aligned to an accuracy of a fraction of a
%pixel. The $R$ band image was chosen as the reference image since it
%has the worst seeing. All other images were registered to the
%reference image using a third order two dimensional fit to the
%positions of 72 stars. The $R$ and $I$ band images were both observed
%with FORS2. A simple shift of the $I$ band image was therefore
%sufficient to match it to the $R$ band image.  The $B$ and narrow band
%image ([\ion{O}{ii}]/8000, called \nbo\ from now on) were observed
%with FORS1.  The plate scale of the FORS1 and FORS2 detectors is
%slightly different: a simple shift of images would result in a
%positional difference of 1\farcs4 in the corner of the CCD.  The $B$
%and \nbo\ images were therefore matched to the reference image by the
%application of a geometric transformation using the above mentioned
%fit.

\begin{table*}
\begin{center}
\caption[]{Observations and filter properties} \label{obs_table}
\begin{tabular}{r@{--} r@{--} r r r r r r r r r} \hline \hline
\multicolumn{3}{l}{Date} & M & Tel/Instr & Filter & $\lambda_{\rm c}$
& $\lambda_{\rm fwhm}$ & Exp & Time & P \\
\multicolumn{3}{c}{(1)} & (2) & \multicolumn{1}{c}{(3)} & (4) &
(5) & (6) & (7) & (8) & (9) \\ \hline
12 & 4 & 1999 & V & UT1/FORS1 & $B$ Bessel & 429  & 88  &  300 & 1800 &1\\
12 & 4 & 1999 & V & UT1/FORS1 & \nbo       & 381.4& 6.5 & 1800 & 9000 &1\\
13 & 4 & 1999 & V & UT1/FORS1 & \nbo       & 381.4& 6.5 & 1800 & 5400 &1\\
6  & 3 & 2001 & S & UT2/FORS2 & $R$ Special& 655  & 165 &  300 & 1800 &1\\
5  & 3 & 2001 & S & UT2/FORS2 & $I$ Bessel & 768  & 138 &  450 & 2700 &1\\
27 & 3 & 2001 & V & UT2/FORS2 & $I$ Bessel & 768  & 138 &  300 & 5400 &1\\
6  & 1 & 2001 & S & UT1/ISAAC & $J_s$      & 1240 & 160 & 45/5 & 3600 &1\\
17 & 4 & 2000 & S & UT1/ISAAC & $H$        & 1650 & 300 & 13/8 & 2496 &1\\
3  & 4 & 1999 & S & UT1/ISAAC & $K_s$      & 2160 & 270 & 10/10& 2900 &1\\
4  & 4 & 1999 & S & UT1/ISAAC & $K_s$      & 2160 & 270 & 10/10& 1900 &1\\
6  & 1 & 2001 & S & UT1/ISAAC & $K_s$      & 2160 & 270 & 13/8 & 2496 &1\\
16 & 4 & 2000 & S & UT1/ISAAC & \nbir      & 2070 & 26  & 60/5 & 5700 &1\\
17 & 4 & 2000 & S & UT1/ISAAC & \nbir      & 2070 & 26  & 60/5 & 6300 &1\\
19 & 4 & 2000 & S & UT1/ISAAC & \nbir      & 2070 & 26  & 60/5 & 5400 &1\\
10 & 1 & 2001 & S & UT1/ISAAC & $K_s$      & 2160 & 270 & 13/8 & 4576 &2\\
11 & 1 & 2001 & S & UT1/ISAAC & \nbir      & 2070 & 26  & 75/5 & 9000 &2\\
7  & 2 & 2001 & S & UT1/ISAAC & \nbir      & 2070 & 26  & 75/5 & 6000 &2\\
21 & 2 & 2001 & S & UT1/ISAAC & \nbir      & 2070 & 26  & 75/5 & 2250 &2\\
\hline \hline
\end{tabular}
\end{center}
%\begin{center}
\footnotesize \noindent Notes: (1) Date: Day -- Month -- Year (2)
Visitor (V) or Service (S) mode (3) Telescope and instrument (4) ESO
filter name (5) Filter central wavelength in nm (6) Filter full width
at half maximum in nm (7) Exposure time for single frames in seconds;
for IR observations DIT/NDIT where DIT is integration time for
sub-integration and NDIT number of sub-integrations (8) Total exposure
time in seconds (9) Pointing on \fone\ or \ftwo\ (see Sect.~\ref{ir_obs}).
%\end{center}
\end{table*}                     

\subsection{Infrared observations}\label{ir_obs}
In 2000 and 2001 Antu's infrared camera ISAAC was employed to carry
out imaging of the field of \rgs\ in several near infrared broad bands
and a narrow band that included the redshifted \ha\ emission line
(\ion{He}{i} filter, \nbir\ from now on). The short wavelength camera
of ISAAC is equipped with a Rockwell Hawaii 1024$^2$ pixel Hg:Cd:Te
array which has a pixel scale of 0.147\arcsec, yielding a field of
$\sim$ 2\farcm5$\times$2\farcm5.  This is significantly smaller than
the field covered by the optical observations, so we carried out two
pointings in the infrared: one centered at the position of the radio
galaxy as the optical observations ($\alpha,\delta_{\rm J2000}$ =
${\rm11^h40^m48^s}$, -26$^\circ$29\arcmin10\arcsec, hereafter \fone)
and one to the North East ($\alpha,\delta_{\rm J2000}$ =
${\rm11^h40^m57^s}$, -26$^\circ$28\arcmin48\arcsec, hereafter \ftwo)
covering six confirmed \lya\ emitters (Paper II).

All infrared images were taken in jitter mode, where the telescope is
offset randomly between exposures but never farther from the original
pointing than 20\arcsec.  In $J_s$, $H$ and $K_s$ individual frames
were exposed for 100 to 225 seconds using sub-integrations of length
10 to 45 seconds to avoid over-exposure of the background. The narrow
band frames were exposed for 300 or 375 seconds with sub-integrations
of 60 to 75 seconds respectively. Specifications (date, mode, band,
integration time and pointing) of all observations are presented in
Table \ref{obs_table}. Note that only \fone\ was observed in $J_s$ and
$H$ band. Observations in $K_s$ of \fone\ were taken in ESO period 63
(P63, 1999) and period 66 (P66, 2001). The sensitivity in P66 had
increased by 45\% compared to P63, amongst others due to an
aluminization of the main mirror.  We have scaled the measurements
done in P63 to P66, effectively reducing the formal exposure time in
P63. The total exposure time in $K_s$ for \fone\ in terms of P66 time
units is 1.6 hours.

The infrared observations were reduced in the same way as the optical
ones. However, the atmospheric emission in the near infrared is
variable on a time scale comparable with the exposure time of
individual frames, causing fringing residuals in the frames after
background subtraction using the median of six to ten frames. These
residuals had to be removed in the $K_s$ and \nbir\ frames observed in
2000 by subtracting a low order polynomial fit to the lines and
columns of the masked images.  An overview of total exposure time,
limiting magnitude and resultant seeing can be found in Table
\ref{reduced}.

The infrared images were registered with the optical reference image
using the same pixel scale. In \fone\ 37 objects were used for the
alignment and in \ftwo\ 40, resulting in less than one pixel
difference between all images over the entire field. The $K_s$ and
\nbir\ images of \fone\ and \ftwo\ were merged into one rectangular
mosaic image.  The overlap in \fone\ and \ftwo\ gives rise to a region
of about one square arcminute in the mosaic where the noise level is
lowest. As a final step a second set of images was made matching the
resolution of the reference image using six stars to estimate the
difference in point spread function.

\begin{table}
\begin{center}
\caption[]{Resultant images} \label{reduced}
\begin{tabular}{l c r c c} \hline \hline
Band & T & Size & M$_{\rm lim}$ & Seeing \\
(1) & (2) & \multicolumn{1}{c}{(3)} & (4) & (5) \\ \hline
\nbo         & 4.0 & 46.6 & 26.5 & 0.75\arcsec\\
$B$          & 0.5 & 46.6 & 27.5 & 0.70\arcsec\\
$R$          & 0.5 & 46.6 & 26.4 & 0.85\arcsec\\
$I$          & 2.0 & 46.6 & 26.8 & 0.65\arcsec\\
$J_s$        & 1.0 &  7.5 & 24.8 & 0.45\arcsec\\
$H$          & 0.7 &  7.5 & 23.8 & 0.70\arcsec\\
$K_s$        & 1.6 & 12.5 & 23.0 & 0.45\arcsec\\
\nbir        & 4.8 & 12.5 & 22.8 & 0.50\arcsec\\
\hline \hline
\end{tabular}
\end{center}
%\begin{center}
\footnotesize \noindent Notes: (1) Broad band or narrow band (see
Table~\ref{obs_table} for specifications) (2) Total exposure time
(hours) (3) Field size in square arcminute (4) 3$\sigma$ limiting Vega
magnitude in 1\arcsec\ aperture as measured on central square
arcminute of image (5) Seeing on resultant image.
%\end{center}
\end{table}

\section{Object detection and photometry}\label{cat}

\subsection{Catalogue sets}
The results of the observations and reduction described above are a
set of FORS 7\arcmin$\times$7\arcmin\ images in \nbo, $B$, $R$ and $I$
and a set of ISAAC 2\farcm5$\times$2\farcm5 images in $J_s$, $H$,
$K_s$ (2$\times$) and \nbir\ (2$\times$). From these, we have created
catalogs of detected and flux-calibrated objects in order to select
(i) objects in $K_s$ band, (ii) EROs, (iii) candidate \ha\ emitters at
$z = 2.16$ and (iv) candidate \lya\ emitters at $z = 2.16$.

We have used the SExtractor software \citep[v2.2.1,][]{ber96} for
object detection and photometry. Since the background noise level
varies across the images as a result of the dithering technique
employed, object detection was not performed directly on the final
reduced images, but on additional images weighted to give a
homogeneous noise level.  These were created by multiplication of the
reduced images by their associated exposure time maps.  Only for the
detection of \lya\ emitters, a homogeneous noise level image convolved
to the $R$ band seeing was used, where the detection sensitivity for
slightly extended objects (0\farcs85) is highest. A disadvantage of
the use of convolved images for object detection is that spurious
sources (e.g.\ remaining cosmic rays) become indistinguishable from
real sources. Three regions in the \lya\ detection image that were
badly affected by bright stars were blanked (for a total of 7.75
arcminute$^2$).  The source extraction parameters were set so that,
detected objects must have at least 8 connected pixels with flux in
excess of 1.5 times the background noise level of the detection image,
except for the \lya\ detection image, where a source has to have 14
connected pixels.  To ensure that the colours are computed correctly,
object photometry was done on the convolved images, by employing
SExtractor's double image mode using the apertures defined on the
weighted images.  A weight map created from the square root of the
exposure time map was used to estimate the errors in the photometry.

\citet{kro80} and \citet{inf87} have shown that for stars and
galaxy profiles convolved with Gaussian seeing, $> 94$\% of the flux
is inside the appropriately scaled Kron aperture.  We have therefore
used SExtractor's MAG\_AUTO implementation of Kron's first moment
algorithm to estimate the \emph{total} magnitudes of the sources.
The resultant magnitudes were corrected for galactic extinction of
A$_{\rm B}$ = 0.172 \citep{sch98} and assuming an R$_{\rm V}$ = 3.1
extinction curve, which resulted in a decrease of the zero-points of
$B$, $R$ and $I$ by 0.2, 0.1 and 0.1 respectively. No changes were
necessary for the infrared zero-points.

%For the Kron magnitude, an elliptical aperture is defined whose
%elongation and position angle are determined from the second order
%moments of the object's light distribution. This ellipse is scaled to
%a size of about two isophotal radii. Within this aperture, the first
%moment radius of the light distribution is computed. The long axis of
%the final aperture in which the flux is determined is equal to a
%scaling factor times the first moment radius times the elongation.  The 
%scaling factor was set to 2.5.

SExtractor classifies the likelihood of detected objects to be
stars or galaxies using a neural network. The resultant
\emph{stellaricity} index has a range from 0.0 to 1.0, where stars
should have a value near 1.0 and galaxies a value near 0.0.

To derive a list of \ha\ emitting candidates, object detection and
aperture definition needed to be done on the infrared narrow band
image. For our first set of catalogues, the apertures were therefore
defined on the unconvolved homogeneous noise level image associated
with the \nbir\ image. In this way, most remaining cosmic rays and CCD
defects are too small to be included in the list and we do not
introduce a preference for a fixed spatial frequency. Photometry with
these apertures was subsequently carried out on all eight convolved
images. The resulting catalogue contains 479 objects, of which
thirteen were either spurious or not suitable for the detection of
line emitters (e.g.\ bright stars, a few remaining cosmic rays and
some image boundary defects).

We are also interested in the population of EROs in the field of
\rg. These objects are old elliptical or dusty starburst galaxies at
$z > 1$ and may also be present in the proto-cluster structure. EROs
have extreme $I - K$ colours, i.e.\ they are detected in $K$ band but
are very faint in the optical. A second set of catalogues was
therefore based on the unconvolved homogeneous noise level image
associated with the $K_s$ band image. From this set, we derive the
$K_s$ band counts and EROs. The resulting catalogue contains 550
objects.

Although candidate \lya\ emitters were selected in Paper I, we have
derived a new list of \lya\ candidates based on the newly reduced $B$
and \nbo\ images using selection criteria consistent with the criteria
for selecting the \ha\ emitters presented in this paper. For this
purpose, a third set of catalogues was constructed, based on the
convolved homogeneous noise level image associated with the \nbo\
image. This set contains 1027 sources.

\section{Number counts and cluster candidates}\label{can}

\subsection{$K_s$ number counts}

$K$ band number counts were derived from the catalogues based on the
$K_s$ image. Table \ref{Kcounts} lists the number of sources per half
magnitude bin and the cumulative number of sources per square
degree. \citet[][B00 from now on]{bes00} shows that selecting $K$ band
objects with SExtractor's stellaricity index below 0.8 efficiently
selects galaxies as opposed to stars. This conclusion is based on the
$J - K$ colour of the detected objects which is in general bluer for
stars. Because \ftwo\ is not imaged in $J$, we determined the galaxy
counts from the total number counts (550 objects) by selecting only
those objects with stellaricity index lower than 0.8 (470 objects).

\begin{table}
\begin{center}
\caption[]{$K_s$ band galaxy counts} \label{Kcounts}
\begin{tabular}{r r r r r r} \hline \hline
$K_s^{\rm lim}$ & n & N$^*$ & 
\multicolumn{1}{r}{N$^*_{\rm Sco}$} & $>$4 & $>$5 \\
(1) & (2) & (3) & (4) & (5) & (6) \\ \hline
18.0 &  19 &  2.5$\pm$0.4 &              &  1 &  0 \\
18.5 &  17 &  3.8$\pm$0.6 &              &  6 &  1 \\
19.0 &  29 &  6.2$\pm$0.7 &  4.3$\pm$0.3 &  9 &  1 \\
19.5 &  34 &  8.9$\pm$0.8 &  6.3$\pm$0.4 & 19 &  2 \\
20.0 &  58 & 13.5$\pm$1.0 &  8.4$\pm$0.4 & 26 &  5 \\
20.5 &  69 & 19.0$\pm$1.2 & 11.3$\pm$0.5 & 37 &  7 \\
21.0 &  59 & 23.8$\pm$1.4 & 16.1$\pm$0.6 & 47 & 11 \\
21.5 &  84 & 30.5$\pm$1.6 &              & 59 & 16 \\
22.0 &  55 & 34.9$\pm$1.7 &              & 64 & 19 \\
\hline \hline
\end{tabular}
\end{center}
%\begin{center}
\footnotesize \noindent Notes: (1) Limiting $K_s$ magnitude (2)
Differential counts between $K_s^{\rm lim} - 0.5$ and $K_s^{\rm lim}$
(3) Cumulative number arcmin$^{-2}$ of galaxies brighter than limiting
$K_s^{\rm lim}$, Poisson error is indicated (4) Same as (3) from blank
field survey by \citet{sco00} (5) Number of objects brighter than
$K_s^{\rm lim}$ with $I - K_s > 4$ (6) Same as (5) for $I - K_s > 5$
(See Sect.\ \ref{ero_sec}).
%\end{center}
\end{table}

The cumulative counts were compared with observations of a blank field
of substantial size (43 arcminute$^2$) by \citet{sco00}.  Although
these authors also make a distinction between stars and galaxies based
on SExtractor's stellaricity index (0.85), it is not clear whether the
number counts in their tables are total counts or galaxies only. We
assume here that galaxy counts are listed. We observe on average
1.5$\pm$0.1 times the number of objects expected from the blank field
survey, as illustrated by the observed surface density of galaxies
brighter than $K_s = 20$ of 13.5$\pm$1.0 arcmin$^{-2}$, compared with
8.4$\pm$0.4 arcmin$^{-2}$ determined by \citeauthor{sco00}. Less deep,
but using a much larger field are the observation of \citet{dad00a},
which determine a $K_s$ galaxy count of 3.29$\pm$0.07 objects
arcmin$^{-2}$ up to $K_s = 18.7$ (which is slightly higher than
\citeauthor{sco00}'s value), while we find 4.5$\pm$0.6 galaxies
arcmin$^{-2}$ up to this limit. From Fig.\ 2 in \citet{dad00a}, it is
clear that there exist considerable scatter in the observations by
different authors, which they interpreted as due to cosmological
field-to-field variations of up to a factor two. Our $K_s$ band counts
are near the upper limit of the observed variations.

\begin{figure}
  \resizebox{\hsize}{!}{\includegraphics{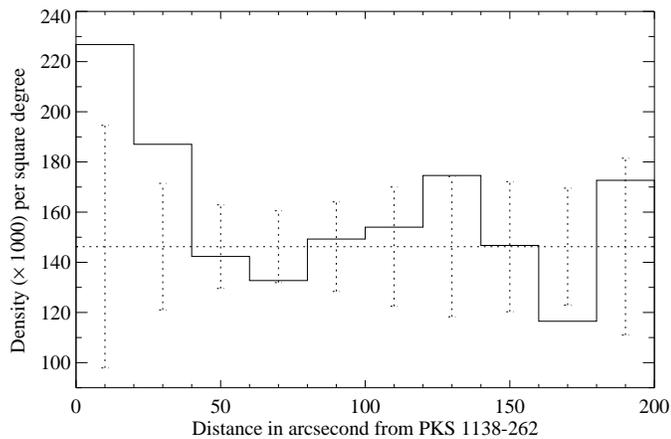}}
  \caption{Galaxy counts as a function of distance from the radio
  galaxy in circular bins of 20\arcsec\ width. The error-bars represent
  Poissonian errors. The mean density (146 $\times 1000$) per square
  degree inside the largest circle (210\arcsec, bin not shown) is
  indicated by the horizontal line.}
  \label{kcountcirc}
\end{figure}

We have analyzed the surface density of $K$ band selected galaxies as
a function of distance from the radio galaxy by counting the number of
galaxies in circular areas around \rgs. Fig.~\ref{kcountcirc} shows
the number of galaxies per arcsecond$^2$ in circular bins of
20\arcsec. It is clear that the counts show an excess within
50\arcsec\ or 0.45 Mpc from the radio galaxy. The deviation from the
mean density of the joined first two bins is 2.2$\sigma$.

%This result is consistent with \rgs\ being located in a cluster sized
%galaxy over-density, although we cannot identify on the basis of number
%counts which objects are part of the structure at $z \sim 2.2$ and
%which are located in the foreground or background. In Sects.\
%\ref{ero_sec} and \ref{ha_em_sec}, we will identify candidate cluster
%members by their red broad band colours or narrow band excess
%emission.

The richness of clusters can be assessed by counting the number of
cluster galaxies found within a radius of 0.5 Mpc of the central
galaxy with magnitudes between m$_1$ and m$_1$+3, m$_1$ being the
magnitude of the central galaxy.  This value, N$_{0.5}$, is defined by
\citet{hil91} and based on an earlier definition by \citet{abe58}. The
$K$ magnitude measured for \rgs\ is 16.1. We can correct this value
for line emission from \ha\ and \ion{N}{ii} as measured in the \nbir\
band, by solving for line and continuum contributions in the broad and
narrow band. We obtain
\begin{equation}c = {b - n \over 
\Delta \lambda_{\rm bb} - \Delta \lambda_{\rm nb}},
\end{equation} 
where $c$, $b$ and $n$ are the continuum flux and flux measured in
broad and narrow band and $\Delta \lambda_{\rm bb}, \Delta
\lambda_{\rm nb}$ are the FWHM of the broad and narrow band,
respectively.  The $K$ magnitude in the \nbir\ band is 15.8 and the
resulting magnitude of the continuum flux therefore 16.2. Note that
this difference of 0.1 units of magnitude is less than the 0.3 units
derived in \citet{pen97} based on an assumed \lya/\ha\ ratio of 1. In
addition to line flux, we have to correct for other non stellar
contributions. \citet{pen97} attribute 0.2 magnitude units to the
point source contribution and a maximum of 0.2 units to the resolved
contribution.  We conclude that the continuum $K$ magnitude
representative of the stellar population in the brightest cluster
galaxy is close to 16.5.  At $z = 2.16$, 0.5 Mpc is equivalent to
56\arcsec, and the angular area occupied by a disc with 0.5 Mpc radius
is 2.7 arcmin$^2$. The number of galaxies between 16.5 and 19.5 in
this area around \rgs\ is 31.  From Table~1 in B00, we find that blank
fields contain $\sim 2.06\times 10^4$ galaxies per square degree
between magnitude 16.5 and 19.5 (references to the blank field data
can be found in the caption of Fig.\ 6 in B00), or 15 galaxies within
a circular area of 56\arcsec\ radius. The net excess count around
\rgs\ is therefore 16$\pm$6. From Table~4 in \citet{hil91} we read
that for clusters of richness 0, 1, 2, the mean N$_{0.5}$ values
determined by \citet{bah81} are 12$\pm$3, 15$\pm$5, 29$\pm$8. The mean
value of N$_{0.5}$ in 3C radio galaxy fields \citep{yat89} is
11$\pm$2, similar to the average number of 11 galaxies in the 28 $z
\sim 1$ radio galaxy fields found by B00. The measured N$_{0.5}$
suggests that \rgs\ is located in an environment with a density
comparable to richness class 0 or 1 clusters. During the evolution of
this structure, however, more galaxies might fall in, further
increasing the richness of the cluster.

\subsection{Extremely red objects}\label{ero_sec}

Since the discovery of extremely red objects
\citep[EROs,][]{els88,els89}, there has been considerable interest for
these high redshift galaxies because their properties can constrain
models of galaxy formation and evolution. They are now generally
believed to be either evolved ellipticals or dusty starbursts at $z >
1$ and have been shown to cluster strongly \citep{dad00a}. Using our
multi band observations, we can search for EROs which could form a
population of galaxies associated with the radio galaxy at $z \sim
2.16$. Clusters are known to possess a population of elliptical
galaxies which form a \emph{red sequence} in a colour-magnitude
diagram \citep[e.g.][]{bow92}. The evolution of the elliptical galaxy
population in clusters has been shown by numerous authors \citep[e.g.\
][]{sta98} to be simple and homogeneous, indicating that the stellar
population that makes up the red sequence is formed at high redshifts
($z_f > 2$). \citet{gla00} show that the red sequence can be exploited
to find clusters of galaxies up to $z \sim 1.4$ using optical
imaging. Basically, the cluster red sequence is as red as or redder
than other galaxies at a given redshift and all lower redshifts if
properly chosen filters straddling the 4000 \AA\ break are used
\citep{gla00}. For a cluster elliptical at $z = 2.2$ the 4000 \AA\
break is redshifted to 12800 \AA, in the infrared $J$ band. We have
selected EROs based on their $I - K_s$ colour, which also targets
galaxies at the redshift of the proto-cluster.  The samples selected
with these bands can be compared with literature data.

\subsubsection{EROs in the field of \rg} \label{sec:eros_field}

\begin{figure}
  \resizebox{\hsize}{!}{\includegraphics{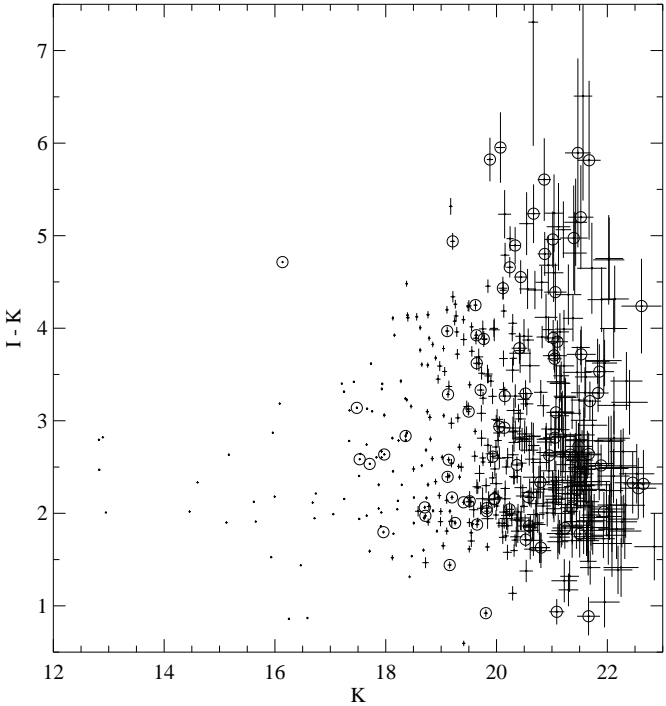}}
  \caption{Colour-magnitude plot of $I - K$ vs $K$ for the 544 sources
  detected on the $K_s$ band image. The sources within 40\arcsec\ of
  the radio galaxy are indicated by circles.}
  \label{kik}
\end{figure}

\begin{figure}
  \resizebox{\hsize}{!}{\includegraphics{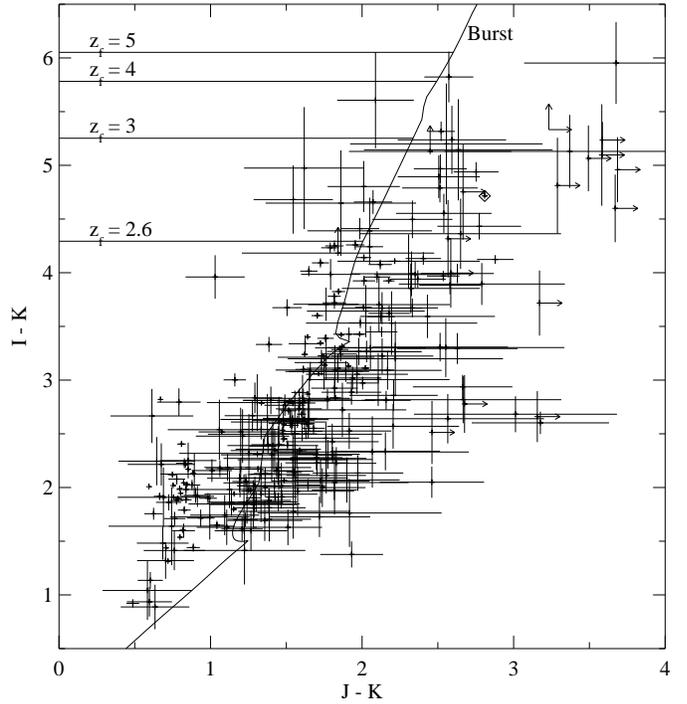}}
  \caption{$I - K$ vs $J - K$ colour-colour plot for the 320 sources
  detected in the central $K_s$ band image. Non-detections in $I$ or
  $J$ and therefore lower limits on the magnitude are indicated by
  arrows. The radio galaxy (indicated by a diamond) has $J - K$ = 2.8
  and $I - K$ = 4.7. The track marks the colour of a stellar
  population at $z = 2.2$ formed by a 100 Myr single burst followed by
  passive evolution for ages of 0.5 Myr to 2.4 Gyr after the
  burst. Four formation redshifts of the population are indicated.
  \rg\ is indicated by a diamond.}
  \label{jkik}
\end{figure}

\begin{figure}
  \resizebox{\hsize}{!}{\includegraphics{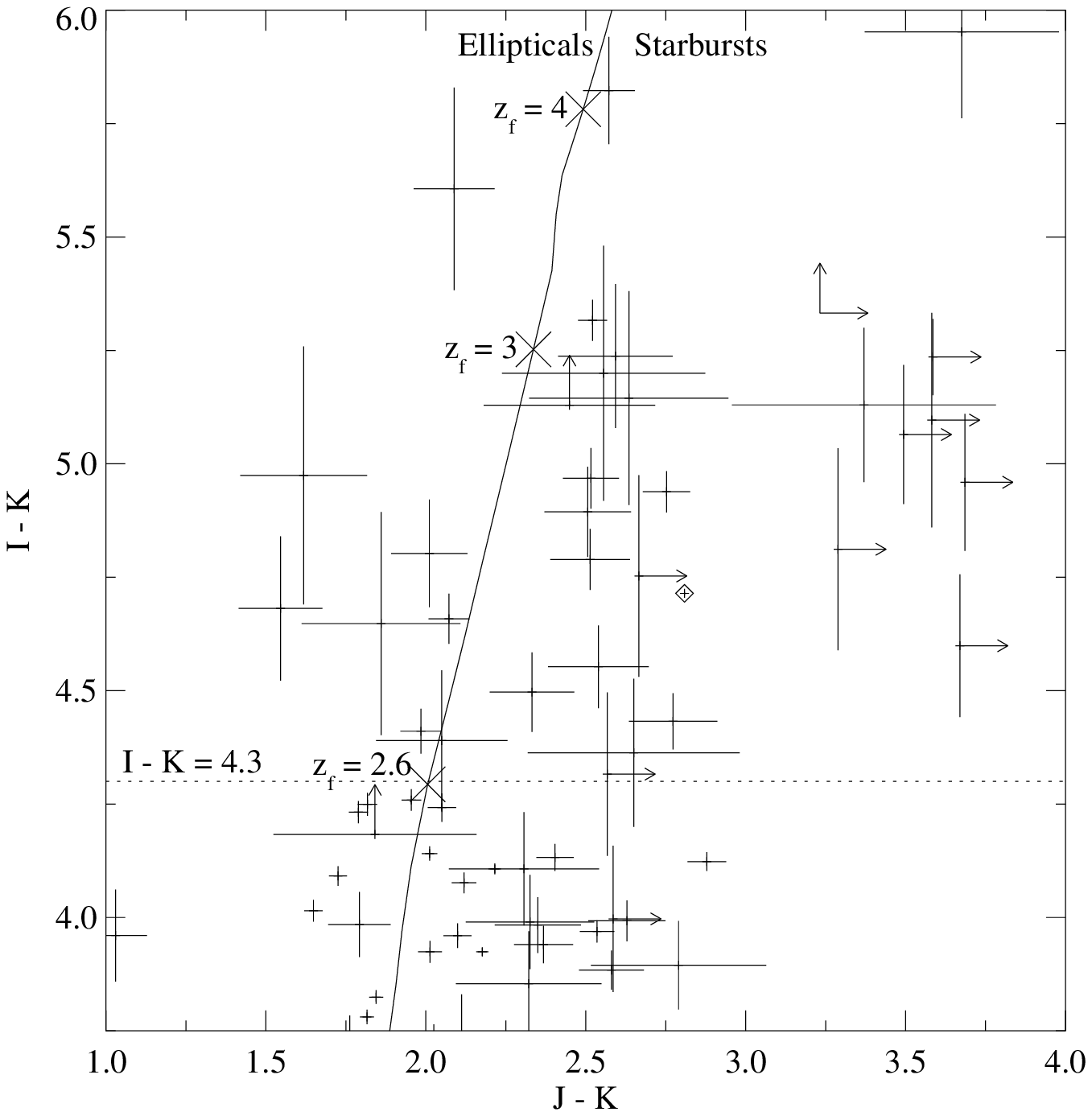}}
  \caption{This close-up of Fig.~\ref{jkik} shows the extremely red
  objects. The horizontal line denotes our ERO selection criterion $I
  - K > 4.3$. Error-bars have been reduced in size by a factor two to
  increase the readability of the plot. The solid line, diamond and
  arrows are described in the caption of Fig.~\ref{jkik}. Three
  formation redshifts for the stellar population are indicated.}
  \label{jkik_ero}
\end{figure}

Figs.\ \ref{kik} and \ref{jkik} show plots of $I - K$ vs $K$ and $I -
K$ vs $J - K$ for the sources with apertures defined on the $K_s$ band
image. These plots show that there is a considerable number of objects
with very red colours ($I - K_s > 4$) in our field. An enlargement of
Fig.\ \ref{jkik} is shown in Fig.\ \ref{jkik_ero} for the range in $I
- K > 3.75$. For $I - K > 4.3$, there are very few bright objects: the
median $K$ band magnitude in the range $4.3 < I-K < 5.1$ is 20.9 while
it is 1.3 magnitude lower in the range $3.5 < I-K < 4.3$, with the
sole exception of the radio galaxy ($I - K = 4.8$, $J - K =
2.8$). This increase in magnitude suggests that a large fraction of
the objects redder than this limit are distant galaxies and we
consider therefore the 44 objects with $I - K > 4.3$ in the field of
\rgs\ as EROs for the remainder of this paper. These objects are
listed in Table~\ref{erotable}. However, since other authors use
different criteria, we have listed the number of red objects according
to several selection criteria in Table~\ref{eros}.

We will compare the number density of EROs in the field of \rgs\ with
the density observed in a 23.5 arcminute$^2$ area of the
\emph{Chandra} Deep Field by \citet{sco00}, which agrees well with the
density of $I - K_s > 4.0$ objects measured by \citet{cow96}. In Table
\ref{Kcounts}, the number of objects with $I-K > 4$ and $> 5$ found in
the field of \rgs\ is shown per $K_s$ magnitude limit. Up to $K_s$ =
21, one expects 33$\pm$4 (6$\pm$2) EROs with $I-K > 4$ (5) in a blank
field of this size, while we observe 47 (11).  Note, however that
\citet{dad00b} claim that the density of EROs with $R - K_s > 5$ and
$K < 19$ in the 43 arcmin$^2$ CDFS, derived by \citet{sco00}, is a
factor of five smaller than the one derived in the 700 arcmin$^2$
survey by \citet{dad00b}. This large discrepancy is not unexpected
given the strong clustering of EROs.  Although we are using different
selection criteria and most EROs detected in our comparatively small
12.5 arcmin$^2$ field are fainter than $K = 19$, we should be careful
drawing conclusions from this comparison. It is unclear whether the
clustering of EROs at fainter flux levels is as strong as at bright
levels, but the measurements by \citet{dad00b} show that the
clustering amplitude of EROs with $K_s < 18.5$ is twice as high as of
EROs with $K_s < 19.2$.  Eqs. 8 and 9 in \citet{dad00b} prescribe the
rms fluctuation of ERO counts due to cosmic variance, given the
clustering strength of EROs. We have made the conservative assumption
that $K = 21$ EROs are clustered as strongly as $K = 19$ EROs,
resulting in an uncertainty on the number of EROs with $I - K > 4$ (5)
of 11 (4).  Therefore, we tentatively find an over-density of about a
factor of 1.5, which might be due to a population of EROs in the
proto-cluster at $z \sim 2.2$ on top of a field population of EROs at
lower redshift.

%\citet{dad00b} argue that the strong clustering of EROs implies
%significant surface density variations even for large observing areas.

%\citet{smi02} suggests that the ERO population is dominated by
%ellipticals at $K < 19.5$ and by dusty starbursts at $K > 19.5$, based
%on the flattening of number counts of EROs fainter than $K \sim 19 -
%20$ in their search for gravitationally lensed EROs. This would
%naturally explain the weaker clustering of faint EROs.
%Based on these results and given that the median $K$ magnitude of the
%EROs in the field of \rgs\ is 20.9, the EROs in this field are
%possibly largely constituted by star forming galaxies.

\begin{table}
\caption[]{Extremely Red Objects counts near \rg} \label{eros}
\begin{center}
\begin{tabular}{c c c } \hline \hline
$I - K$ & \# & \#C \\
(1) & (2) & (3) \\ \hline
4.3 - 5.3  & 36 & 27 \\
$>$ 5.3    &  8 &  6 \\
4.5 -- 5.5 & 25 & 22 \\
$>$ 5.5    &  7 &  5 \\
\hline \hline
\end{tabular}
\end{center}
%\begin{center}
\footnotesize \noindent Notes: (1) $I -K$ colour (2) Number of EROs in
12.5 arcminute$^2$ field covered by both K band images (3) Number
of EROs in 7.1 arcminute$^2$ field covered by central $K_s$ band
image only.
%\end{center}
\end{table}

\begin{figure}
  \resizebox{\hsize}{!}{\includegraphics{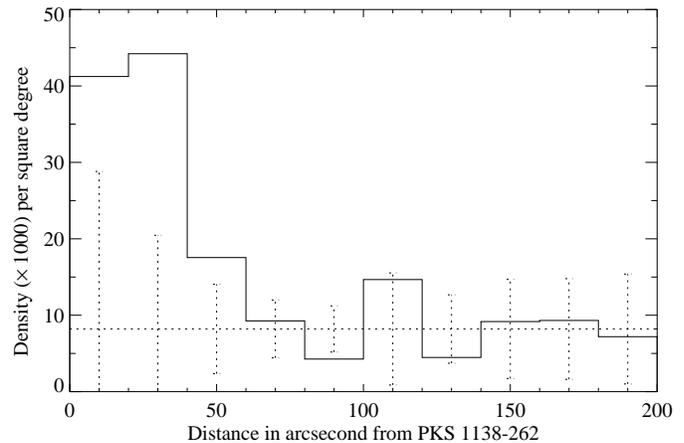}}
  \caption{Extremely red object ($I - K > 4.3$) counts as a function
  of distance from the radio galaxy. The EROs are counted in circular
  bins of 20\arcsec\ width. The error-bars represent Poissonian
  errors. The mean density (8 $\times 1000$) per square degree outside
  60\arcsec\ (0.5 Mpc) is indicated by the horizontal line.}
  \label{ero_circ}
\end{figure}

An important argument for the proposition that part of the ERO
population consists of proto-cluster members at $z = 2.2$ is the
gradient in the spatial distribution of EROs, as shown in Fig.\
\ref{ero_circ}, which is similar to the distribution of $K_s$ band
counts. The density within a 40\arcsec\ radius of the radio galaxy is
more than four times higher than the mean density outside a 60\arcsec\
(0.5 Mpc) radius. This mean surface density of 8$\times 10^3$ galaxies
per square degree would imply a number of 28 EROs in the field,
roughly consistent with the number expected from the blank field
observed by \citet{sco00}. In Fig.~\ref{kik}, the objects within
40\arcsec\ from the radio galaxy are indicated by circles. It is clear
that many of the reddest objects lie near the radio galaxy: there are
17 EROs within a 40\arcsec\ radius of \rgs\ (12 arcmin$^{-2}$) and 27
outside this radius (2.4 arcmin$^{-2}$). The over-density of EROs in
the field of \rgs\ is therefore mostly due to the red galaxies near
the radio galaxy, which is consistent with the observed excess of EROs
being due to a cluster population associated with the radio galaxy.

\subsubsection{EROs in the proto-cluster at $z = 2.2$}\label{sec:eros_proto}

What colours do we expect for galaxies at $z = 2.2$? We have computed
evolutionary tracks for several stellar population using the Galaxy
Isochrone Synthesis Spectral Evolution Library
\citep[GISSEL93,][]{bru93}. The IMF we have used in the models is a
\citet{sal55} law with lower mass cutoff at 0.1 M$_\odot$ and upper
mass cutoff at 125 M$_\odot$. On Fig.~\ref{jkik} a track is indicated
for a stellar population at $z = 2.2$ formed by passive evolution
after a 100 Myr single burst model. The track starts at an age of 5
Myr at $I - K = 0.5$ and leaves the plot at an age of 2.4 Gyr ($I - K
= 6.5$). We have also computed the colours for a constant star
formation model, and an exponential star formation model with $\tau =
1$ Gyr, but these did not reach $I - K > 4$ even after 4 Gyr, the
maximum age for a galaxy at $z = 2.2$ in this cosmology. The EROs in
the field of \rgs\ have colours consistent with galaxies which have
undergone a starburst 0.5 Gyr ago ($z = 2.6$) or longer and evolved
passively thereafter. However, dust reddening is not taken into
account in the models presented here. Reddening of actively star
forming galaxies by dust can also shift their $I - K$ colours into
this regime as shown by e.g.\ \citet{poz00}. The $I - K$ colour of
their elliptical template increases to $\sim 7$ at $z > 2$, while the
colour of their dusty starburst template stays constant at $\sim
6$. We observe no objects with $I - K \sim 7$, but there are three
objects with lower limits to their $I - K$ colour of $\sim 6$, which
could be $z \sim 2$ ellipticals and a few objects with $I - K \sim 6$.
We conclude that the excess of $\sim$ 10 -- 15 of the EROs in the
field of \rgs\ is caused mostly by galaxies at $z = 2.2$ which are
reddened by dust.

%According to \citet{poz00} the age and dust degeneracy of EROs can be
%resolved by their $J - K$ vs $I - K$ colour. Old ellipticals at $1 < z
%< 2$ are brighter in $J$ than dusty starbursts and therefore on the
%left of the dividing line $(J - K) = 0.36 (I -K) + 0.46$. This line is
%not drawn on Fig. \ref{jkik_ero} but is at the same location as the
%track of a passively evolving stellar population indicated by the
%solid line. Note, however, that submillimeter observations of one of
%the reddest galaxies known \citep[object 10 in][]{hu94} show that this
%galaxy at $z = 1.44$ experiences vigorous star formation
%\citep[$\dot{\rm M} \sim 1000$ M$_\odot$ yr$^{-1}$,][]{cim98,dey99}.
%The colours of this object, $I - K = 6.0$ and $J - K = 2.6$
%\citep{gra96}, places it right on top of the line which should
%separate passive from active galaxies. Fig.\ \ref{jkik_ero} shows that
%of the 33 EROs with $I - K > 4.3$ in the central field, 7 are
%designated as ellipticals and 26 as starbursts (including the radio
%galaxy). For galaxies at redshift $z \geq 2$, the 4000 \AA\ break
%moves into the rest frame $J$ band and it becomes impossible to
%determine the nature of EROs on the basis of these colours.

\subsection{Candidate H$\alpha$ emitters}\label{ha_em_sec}

\subsubsection{Selection procedure}\label{haselection}

\begin{figure}
  \resizebox{\hsize}{!}{\includegraphics{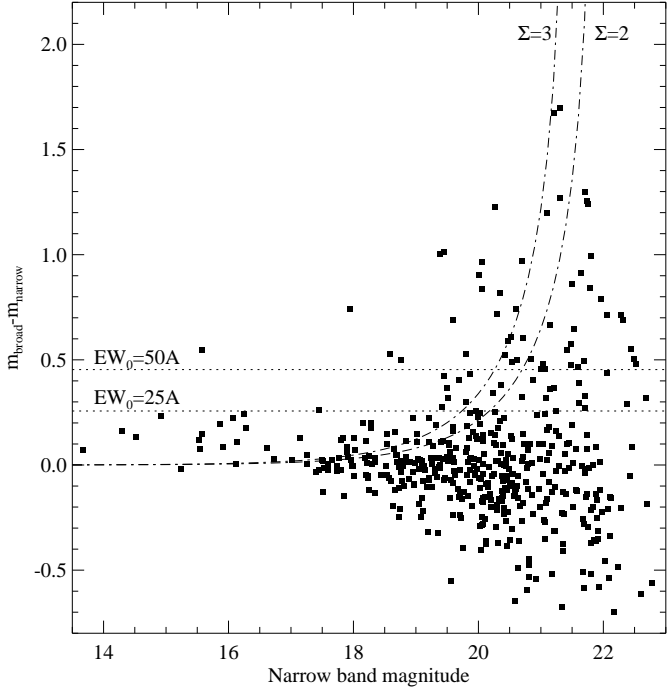}}
  \caption{Colour-magnitude diagram for 467 sources detected in the
  NB~2.07 image. The dot-dashed lines are lines of constant excess
  signal $\Sigma$ (see text). Also shown are lines of constant EW$_0$
  (for $z=2.16$). Candidate \ha\ emitters are objects with EW$_0 >$ 25
  or 50 \AA\ and $\Sigma >$ 2 or 3 (see table \ref{selection}).}
  \label{ha_selection}
\end{figure}

Candidate line emitting objects were selected on the basis of their
excess narrow versus broad band flux, following the criteria of
\citet{bun95} and \citet{moo00}. The selected candidates fulfill two
criteria: first, they have sufficient equivalent width (EW) and
second, their broad band flux is significantly lower than expected for
a flat spectrum source. Having measured the narrow band flux for each
source, we compute the expected broad band flux and its standard
deviation assuming a flat spectrum. The error parameter $\Sigma$ is
defined as the number of standard deviations the measured broad band
flux deviates from the expected broad band flux of a flat spectrum
source \citep[see also][]{bun95}. Note that $\Sigma$ is well defined
for objects not detected in the broad band. A NB~2.07 $ - K_s$ band
versus NB~2.07 magnitude plot for the 466 bona fide objects in the
\ha\ selection catalog (see Sect.\ \ref{cat}) is shown in Fig.\
\ref{ha_selection}.  Also drawn are two horizontal lines indicating
rest frame equivalent width (EW$_0$) of 25 and 50 \AA\ and two curves
indicating $\Sigma$ equal to 3 and 2. The curves of constant $\Sigma$
have been computed for median narrow band and broad band errors; the
actual $\Sigma$ of individual sources depends, amongst others, on the
aperture size and local background noise.

We find 17 candidate emitters with rest-frame equivalent width EW$_0 >
50$ \AA\ and $\Sigma > 3$, all of which have narrow band magnitudes
$\leq 21.3$.  One of these objects is the radio galaxy, while a second
is within the extent of the \lya\ halo of the radio galaxy. If we
lower the selection criteria to EW$_0 > 25$ \AA\ and $\Sigma > 2$, we
find 40 candidates. All of these have m$_{\rm NB\,2.07} \le 21.6$. In
addition to the two objects identified above, there is one more object
in this list within the radio galaxy \lya\ halo. The \ha\ halo of the
radio galaxy is interesting in itself, especially in comparison with
the \lya\ halo and will be studied in another article (Kurk et al.\ in
preparation). Table \ref{selection} lists the number of candidates in
both fields for several selection criteria, while Table
\ref{Hacandidates} lists the $K$ magnitude and emission line
properties, for the candidates with EW$_0 >$ 25 \AA\ and $\Sigma
>$ 2. The NB~2.07 narrow band filter used also includes the
[\ion{N}{II}]$\lambda\lambda 6548,6584$\AA\ lines at $z=2.16$, but in
what follows, we will refer to the combined \ha\ + \ion{N}{II} flux
and equivalent width, as \ha\ flux and equivalent width respectively,
unless otherwise noted.

\begin{table}
\begin{center}
\caption[]{Properties of the samples of \ha\ candidates} \label{selection}
\begin{tabular}{c c c c c c c c c} \hline \hline
EW$_0$ & $\Sigma$ & m$_{\rm NB}$ & \# & \#H & n$_{1138}$ & n \\
(1) &(2)&(3) &(4)&         (5) &  (6) &  (7) \\ \hline
 50 & 3 & --   & 17 & 2 & 1.4$\pm$0.3 \\
 25 & 3 & --   & 23 & 3 & 1.8$\pm$0.4 \\
 25 & 2 & --   & 40 & 3 & 3.2$\pm$0.5 \\
 50 & 1 & --   & 48 & 2 & 3.8$\pm$0.6 \\
 25 & 1 & --   & 60 & 3 & 4.8$\pm$0.6 \\
 75 & 2 & 19.3 &  1 & 0 & 0.1$\pm$0.1 & 0.61$\pm0.35^a$ \\
100 & 3 & 19.5 &  2 & 1 & 0.2$\pm$0.1 & 0.12$\pm0.05^b$ \\
\hline \hline
\end{tabular}
\end{center}
%\begin{center}
\footnotesize \noindent Notes: (1) Rest frame equivalent width lower
limit (2) Signal to noise lower limit \citep[as defined by][]{bun95}
(3) Additional selection criterion: narrow band magnitude upper limit
(4) Number of candidates in 12.5 arcminute$^2$ area (5) Number of
candidates within the radio galaxy \lya\ halo (6) Surface density of
candidates in the field of \rgs\ (arcmin$^{-2}$) (7) Surface density
in other fields: $^a$ \citet{bun95}, $^b$ \citet{wer00}.
%\end{center}
\end{table}                     

\subsubsection{Number density of candidate \ha\ emitters}
In recent years the number of successful searches for \ha\ emitters in
the infrared has grown. \citet{bun95} have obtained near infrared
narrow band images in a 4.9 arcminute$^2$ field towards the quasar
PHL~957, in an attempt to detect \ha\ emission from a damped \lya\
absorber at $z = 2.313$. They find 3 candidate \ha\ emitters at $2.29
< z < 2.35$ with EW$_0 >$ 75 \AA\ and $\Sigma >$ 2, brighter than a
narrow band magnitude limit of 19.3. More recently, \citet{wer00}
presented a survey for \ha\ emission at redshifts from 2.1 to
2.4. They have observed several fields containing known damped \lya\
systems (also including the field of PHL~957) and radio galaxies as
well as random fields for a total area of 55.9 arcminute$^2$. They
detect two radio galaxies and a damped \lya\ absorber in the field of
PHL~957 in \ha\ emission accompanied by two close emitters which had
not been observed before. In addition, they find another candidate in
the field of a radio galaxy, but at a large distance from it
(81\arcsec), which seems to be a merging galaxy. In total, they find
seven \ha\ emitters with EW$_0 >$ 100 \AA\ and $\Sigma >$ 3 down to an
area weighted narrow band magnitude limit of 19.5. The narrow band
fluxes of these candidates are $> 2.0$ $\times 10^{-16}$ \ecs.

Our VLT near infrared imaging is much deeper and most of our
candidates have fluxes below this limit. To compare the number density
of \ha\ emitters in our field with the above mentioned authors, we
have done the selection according to their limits, including the lower
narrow band magnitude limits imposed by their shallower observations.
We find only one and two candidates (see Table \ref{selection}),
resulting in number densities (with large Poisson errors) slightly
lower and higher than \citeauthor{bun95}'s and \citeauthor{wer00}'s
surveys, respectively.

%Using the \citet{bun95} criteria, we find 2 candidates (including one
%located in the radio galaxy halo). Considering our field size of 12.5
%arcminutes, this number is within Poisson errors consistent with the
%3 candidates found by \citet{bun95}. If we select using the the same
%criteria as \citet{wer00}, we find two candidates (one is located in
%the radio galaxy halo). Since our field is 4.5 times smaller than
%\citet{wer00}'s, the number density in our field is about two times
%higher, but due to the small number statistics, they could also be
%equally rich.

%Although the current statistics are not sufficient, the fact that the
%number density of \ha\ candidates in our field is comparable to or
%higher than the density of other targeted fields, supports the idea of
%an over-density of galaxies around \rgs.

\subsubsection{Spatial distribution}

\begin{figure}
  \resizebox{\hsize}{!}{\includegraphics{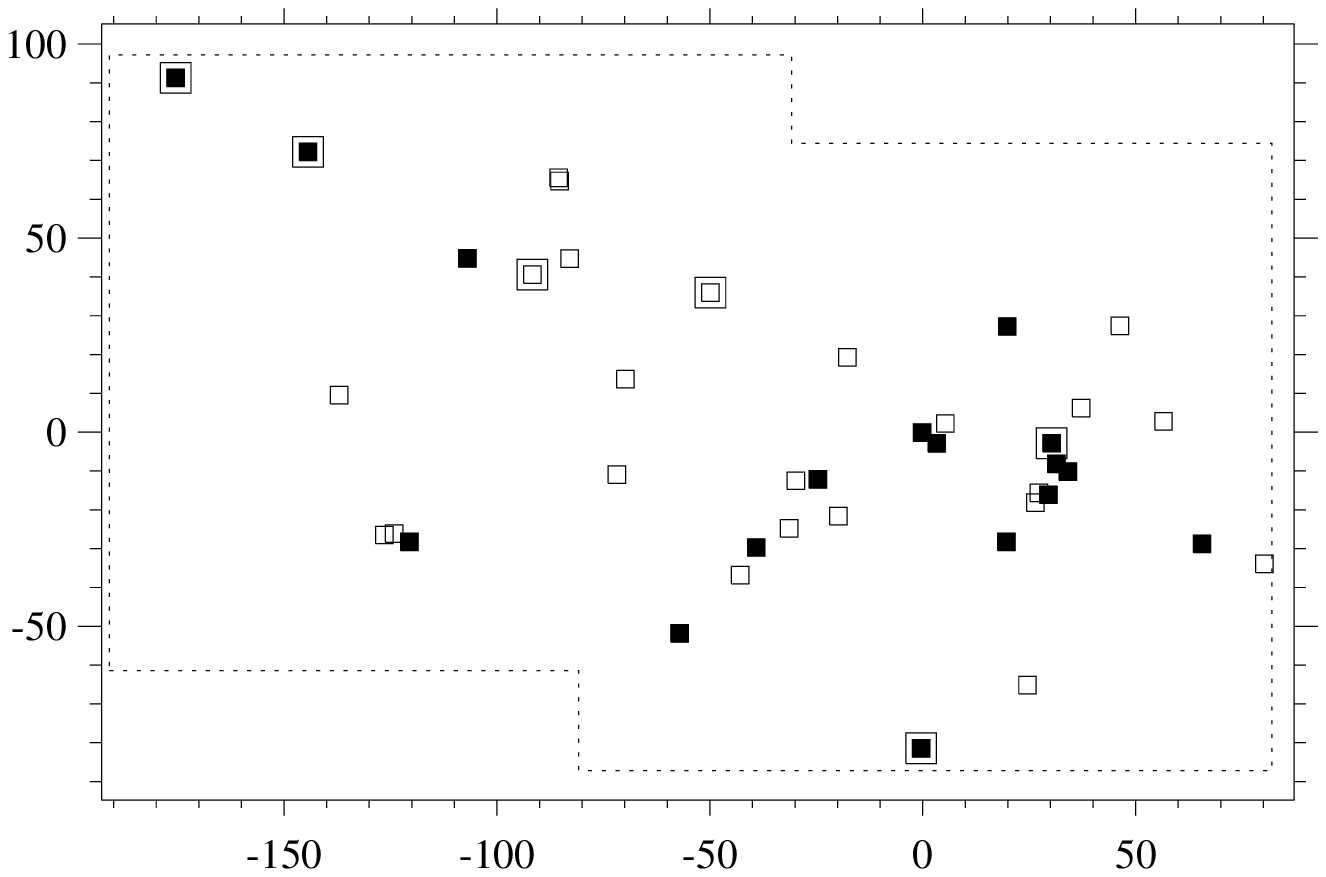}} \caption{Position
  plot for the 40 (17) \ha\ candidates with EW$_0 >$ 25 (50) \AA\ and
  $\Sigma >$ 2 (3) indicated by open (filled) squares.  Large
  squares indicate candidates for which \lya\ emission has been
  detected (see Sect.\ \ref{lyaha_ratio}). Axes are in arcseconds, the
  radio galaxy is at the origin. The dotted boxes indicate the borders
  of the two reduced ISAAC fields.}  \label{ha_pos}
\end{figure}

\begin{figure}
  \resizebox{\hsize}{!}{\includegraphics{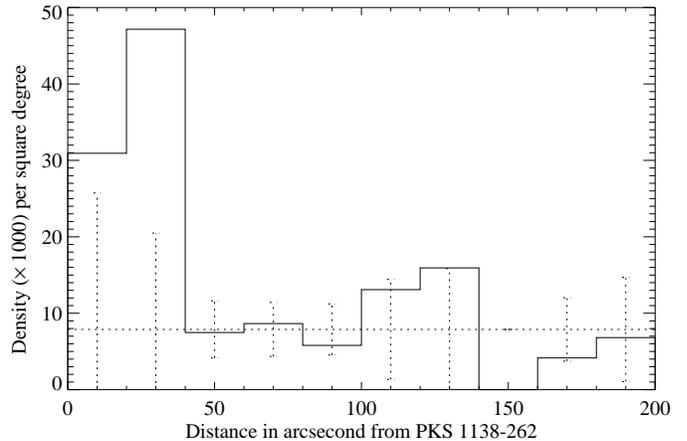}}
  \caption{Candidate \ha\ emitter (EW$_0 >$ 25 \AA, $\Sigma >$ 2)
  counts as a function of distance from the radio galaxy. The
  candidates are counted in circular bins of 20\arcsec\ width. The
  error-bars represent Poissonian errors. The mean density (7.9
  $\times 1000$) per square degree outside 1 arcminute (0.5 Mpc) is
  indicated by the horizontal line.}  \label{ha_circ}
\end{figure}

Fig.\ \ref{ha_pos} shows the spatial distribution of candidate \ha\
emitters.  This distribution is not homogeneous over the observed
fields. To quantify this inhomogeneity, we have counted the number of
candidates with EW$_0 > 25$ \AA\ and $\Sigma > 2$ in circular bins of
20\arcsec\ radius around the radio galaxy. We have taken into account
the variation in sensitivity per pixel using the weight map associated
with the NB~2.07 image. This alters the density per bin by less than
20\% in all bins compared with the unweighted computation. The \ha\
candidate selection depends also weakly on the $K_s$ band sensitivity,
but we do not expect this to have a large influence on the density per
bin and certainly not on the conclusions from this plot. The density
within a 40\arcsec\ distance of the radio galaxy is 5.0$\pm$0.9 times
higher than the mean density outside a 60\arcsec\ (0.5 Mpc)
radius. The high number of candidates near the radio galaxy is
consistent with \rgs\ being located in a region that is over-dense in
\ha\ emitting galaxies.

%The deviation from the mean density of the joined first two bins is
%2.9$\sigma$.

\subsubsection{Contamination by other line emitters}

We do not expect that a significant fraction of our emission line
candidates are line emitters at lower redshift, since there are no
strong emission line redward of \ha\ at 6563 \AA\ (Pa$\alpha$ with a
strength $\sim 0.2$ times the strength \ha\ would imply $z = 0.1$). It
is possible that we have selected galaxies at $z \sim 3.13$ with
strong [\ion{O}{iii}]$\lambda5007$ \AA\ emission or galaxies at $z
\sim 4.55$ with [\ion{O}{ii}]$\lambda3727$ \AA\ emission. The
luminosity distance for objects at $z \sim 4.6$ is 2.5 times larger
than for objects at $z \sim 2.16$. If any of the emitters is indeed an
[\ion{O}{ii}] emitter at this redshift, its intrinsic luminosity is
six times higher than that assumed for \ha\ in Table
\ref{Hacandidates}.  It is therefore unlikely that these objects
contaminate our sample significantly. More likely is the contribution
of [\ion{O}{iii}] emitters to our sample, illustrated by follow-up
spectroscopy to narrow band observations similar to ours.  All
six galaxies suspected to be \ha\ emitters at $z \sim 2.2$ from low
signal-to-noise near-infrared spectroscopy by \citet{moo00} were found
to be [\ion{O}{iii}] emitters at higher redshift when observed at high
enough signal-to-noise to detect the [\ion{O}{iii}]$\lambda4959$ \AA\
line (P.\ van der Werf, private communication). Although on the basis
of the present data, we cannot separate \ha\ emitters at $z \sim 2.2$
from line emitters at higher redshift, the spatial distribution of the
candidate \ha\ emitters indicates that at least $\sim 12$ are
associated with the radio galaxy and are therefore bona fide \ha\
emitters surrounding the cluster.  There is, however, one candidate
which stands out by its brightness ($K = 17.7$) and its size ($\sim$
3\arcsec) from the others ($K \geq 19.3$). The narrow band excess
emission from this candidate is most likely caused by Pa$\alpha$
emission at 1.875\,$\mu$m redshifted to $z = 0.104$ and marked as
such in Table \ref{Hacandidates}.

\subsection{\lya\ candidates}

\subsubsection{Selection procedure}\label{lyaselection}

\begin{figure}
  \resizebox{\hsize}{!}{\includegraphics{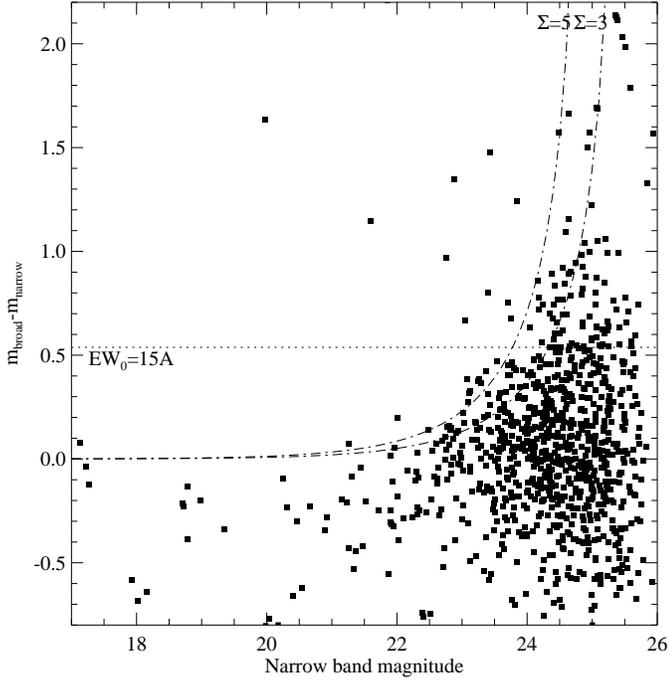}}
  \caption{Colour-magnitude diagram for 1018 sources detected in the
  \nbo\ image. The dot-dashed lines are lines of constant $\Sigma$
  (see text). Also shown is a line of constant rest frame equivalent
  width (for $z=2.16$). Candidate \lya\ emitters are objects with
  EW$_0 >$ 15 \AA\ and $\Sigma >$ 3 or 5.}
  \label{lya_selection}
\end{figure}

Using the catalogue with apertures based on the convolved and partly
blanked \nbo\ image, candidate \lya\ emitting objects were selected on
the basis of excess narrow versus broad band fluxes, similar to the
method used to select \ha\ candidate emitters. Seven spurious objects
near the edges of the images, one saturated star and one blended
object were removed from the catalogue. The difference between the
narrow and broad band magnitude against the narrow band magnitude for
the 1018 definite objects is plotted in Fig.\ \ref{lya_selection}. A
horizontal line indicates rest frame equivalent width (EW$_0$) of 15
\AA\ and curves are drawn for $\Sigma$ equal to 5 and 3. The curves of
constant $\Sigma$ have been computed for median narrow band and broad
band errors; the actual $\Sigma$ of individual sources depends,
amongst others, on the aperture size and local background noise. We
find 11 candidate emitters with rest-frame equivalent width EW$_0 >
15$ \AA\ and $\Sigma > 5$. If we lower the signal-to-noise criterion
to $\Sigma > 3$, we find 40 candidates (see Table
\ref{Lyacandidates}).  One of these objects is the radio galaxy, while
two more are also within the extent of the \lya\ halo of the radio
galaxy. We have checked the number of candidates which would be
selected out of the new catalogues using the criteria used in Paper I
(EW$_\lambda > 65$ \AA\ and $\mathcal{F}_{\rm\nbo} > 2 \times
10^{-19}$ \ecs \AA$^{-1}$). We find 70 candidates, which is consistent
with the 60 candidates found in Paper I, given that the image on which
the current selection is done is $\sim$ 24\% larger than the image on
which the catalogue of Paper I was based.

%Note that of the fifteen \lya\ emitters confirmed in Paper II, twelve
%are also between these candidates. Two are not detected by SExtractor
%due to their low surface brightness in the resampled \nbo\ image and
%one has specifications below the selection criteria (EW$_0 = 11$ \AA),
%but this serendipitous object was also not in the original selection
%catalogue Paper I.

\subsubsection{Spatial distribution}

\begin{figure}
  \resizebox{\hsize}{!}{\includegraphics{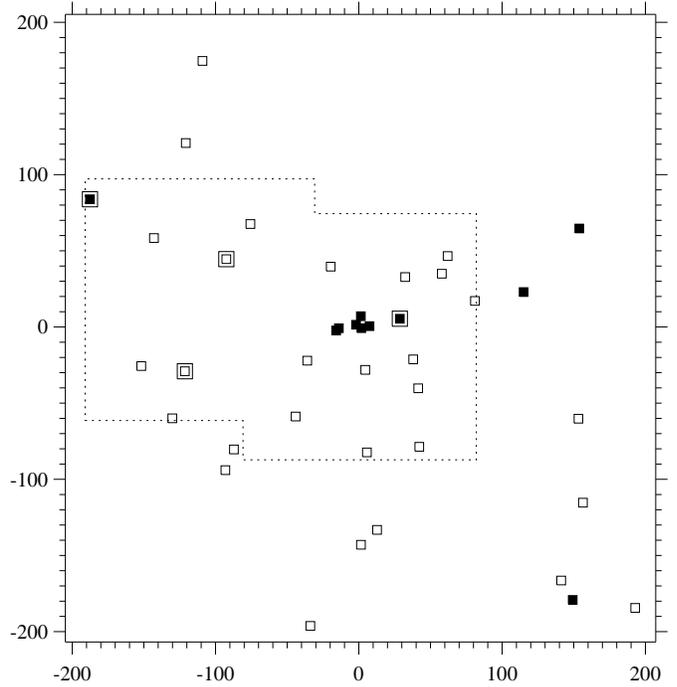}}
  \caption{Position plot for the 40 (11) \lya\ candidates with EW$_0
  >$ 15 \AA\ and $\Sigma >$ 3 (5) indicated by open (filled)
  squares.  Large squares indicate candidates for which \ha\
  emission has been detected (see Sect.\ \ref{lyaha_ratio}).  Axes are
  in arcseconds, the radio galaxy is at the origin. The dotted boxes
  indicate the borders of the two reduced ISAAC fields.}
  \label{lya_pos}
\end{figure}

\begin{figure}
  \resizebox{\hsize}{!}{\includegraphics{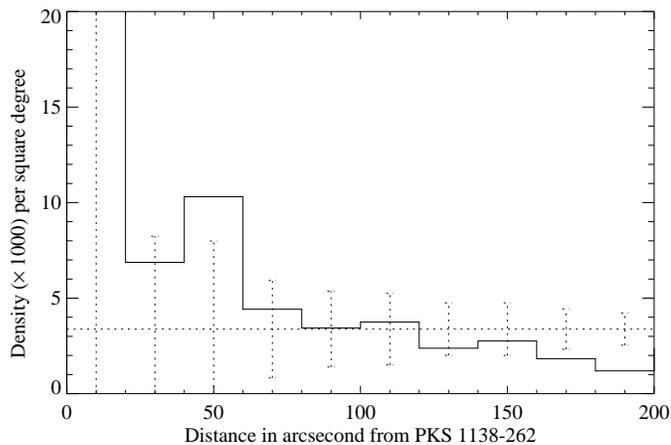}}
  \caption{Candidate \lya\ emitter (EW$_0 >$ 15 \AA, $\Sigma >$ 3)
  counts as a function of distance from the radio galaxy. The
  candidates are counted in circular bins of 20\arcsec\ width. The
  error-bars represent Poissonian errors. The mean density (3.4
  $\times 1000$) per square degree is indicated by the horizontal
  line.}  \label{lya_circ}
\end{figure}

Fig.\ \ref{lya_pos} shows the spatial distribution of the \lya\
candidate emitters. In the south east corner of the image is a bright
star and a nearby galaxy which inhibits the detection of any faint
\lya\ emitters. Although the distribution of the candidates is not
homogeneous, there is not a strong indication of a density
concentration within 40\arcsec\ of the radio galaxy, but at distances
$>$ 120\arcsec\ the density is somewhat below the mean, as Fig.\
\ref{lya_circ} shows. The first bin in this plot contains six objects:
the radio galaxy, three objects which are part of the filamentary
\lya\ halo and two more which might be associated to the halo. Most
other bins are consistent with the mean density of 3.4 $\times 1000$
candidates per degree$^2$.

\subsubsection{Number density of candidate \lya\ emitters}
To determine the galaxy over-density near \rg, we would like to
compare the number density of \lya\ emitters found in our field to the
number density of \lya\ emitters in a blank field. In recent years, a
number of surveys for \lya\ emitting objects at high redshift have or
are being carried out. The utility of \lya\ selected galaxies,
representing the faint end of the galaxy luminosity function, as
tracers of large scale structure was illustrated by \citet{fyn01}. The
redshifts and positions of eight objects at $z \sim 3.0$ found by
their \lya\ emission \citep{fyn00} are consistent with this structure
being a single string spanning about 5 Mpc, the first signal observed
of a filament at high redshift \citep{mol01}. Larger surveys have been
carried out at higher redshifts: 3.4 \citep{hu98}, 4.5
\citep{hu98,rho00}, 4.9 \citep{ouc02} and 5.7 \citep{rho01}. One of
the notable results of these surveys is that the observed equivalent
widths of the \lya\ emitters indicate that they are young galaxies
undergoing their first burst of star formation. There are currently no
observations available of \lya\ emitters at $z \sim 2.2$ in a blank
(or any other) field, although such a survey is underway at the Nordic
Optical Telescope which has the necessary but rare high UV throughput
\citep[e.g.\ ][]{fyn02}. We will therefore compare our results to the
observations of a known over-density of LBGs at $z = 3.09$ imaged in
\lya\ \citep[][S00 from now on]{ste00} and a shallower but much larger
survey at $z = 2.42$ \citep{sti01}, assuming that the results obtained
at higher redshift are also applicable at $z = 2.2$. This assumption
seems to be justified by the conclusion of \citet{yan02} that the
galaxy luminosity function does not evolve significantly from $z \sim
3$ to $z \sim 6$ and by the observation of \citet{ouc02} that the
\lya\ and UV-continuum luminosity functions of \lya\ emitters show
little evolution between $z = 3.4$ and $z = 4.9$.

S00 have selected a sample of 72 bona fide line excess emitters with a
narrow band magnitude limit of NB$_{\rm AB}$ = 25.0 and observed frame
equivalent width (EW$_\lambda) > 80$ \AA. This magnitude limit
corresponds to \nbo\ = 24.3. The number of \lya\ candidates in the
effective 43.6 arcminute$^2$ field of \rgs\ with EW$_\lambda > 80$
\AA\ and \nbo\ $< 24.3$ is 11. The surface density of these objects is
0.25 arcmin$^{-2}$, about a fourth of the value (0.96 arcmin$^{-2}$)
measured by S00. Taking into account the redshift range of detectable
\lya\ emitters corresponding to the FWHM of the \nbo\ filter gives a
comoving volume density of 0.0011 Mpc$^{-3}$. The 72 candidates
detected by S00 are located in a comoving volume of 21041 Mpc$^3$ in
our cosmology, resulting in a volume density of 0.0034 Mpc$^{-3}$. The
overdensity of galaxies at $z = 3.09$ is a factor six, consistently
determined by S00 from the redshift density of LBGs as compared with
the general LBG redshift distribution and from the comoving volume
density of the \lya\ candidates at $z = 3.09$ compared to a blank
field survey at $z = 3.43$ \citep{cow98}. Since the comoving volume
density of candidate \lya\ emitters in our field is 3.1 times smaller
than the density found by S00, we estimate the galaxy overdensity in
the field of \rg\ to be a factor 2$\pm$1.  A more direct comparison
with \citet{cow98} also gives a volume overdensity of 1.6$\pm$0.7.
The quoted errors are derived from the Poisson noise on the number of
emitters in the three fields, but the uncertainty due to the
difference in blank field number density at $z \sim 2.2$ and $z \sim
3.1, 3.4$ might be larger.

\citet{sti01} have carried out a search for \lya\ emitters in a field
of 1200 square arcminutes using a medium band filter and found 58
candidates at $z = 2.4$. Since this sample has on average a red colour
($<B - I> = 1.8$), they conclude that the emitters contain an older
stellar component and have therefore undergone their major episode of
star formation at higher redshift. The 58 candidates have continuum
subtracted narrow band fluxes $> 2.0 \times 10^{-16}$ \ecs. Taking
into account the difference in luminosity distance, we find 3
candidates with fluxes $> 2.6 \times 10^{-16}$ \ecs\ in the field of
\rgs. This amounts to a surface density of 0.07 candidates
arcmin$^{-2}$, about a factor two more than the 0.048 sources
arcmin$^{-2}$ found by \citeauthor{sti01} ~The 4\% filter used in
their survey implies a comoving volume of $7.6 \times 10^5$ Mpc$^3$
and a comoving volume density of $7.6 \times 10^{-5}$ Mpc$^{-3}$,
while the 8 candidates near \rgs\ yield a comoving volume density of
$3.1 \times 10^{-4}$ Mpc$^{-3}$. The overdensity implied by the
difference in comoving volume density is a factor 4$\pm$2. Although we
have not corrected our sample of candidate \lya\ emitters for low
redshift interlopers, it is evident that the field of \rgs\ contains
an overdensity of emitters with respect to blank field. In Sect.\
\ref{clustermass} we discuss the overdensity of the \lya\ emitters
with confirmed redshifts and estimate the mass of the proto-cluster
implied.

\subsection{Coincidence of \lya\ and \ha\ emitters and EROs}

\begin{figure*}
  \resizebox{\hsize}{!}{\includegraphics{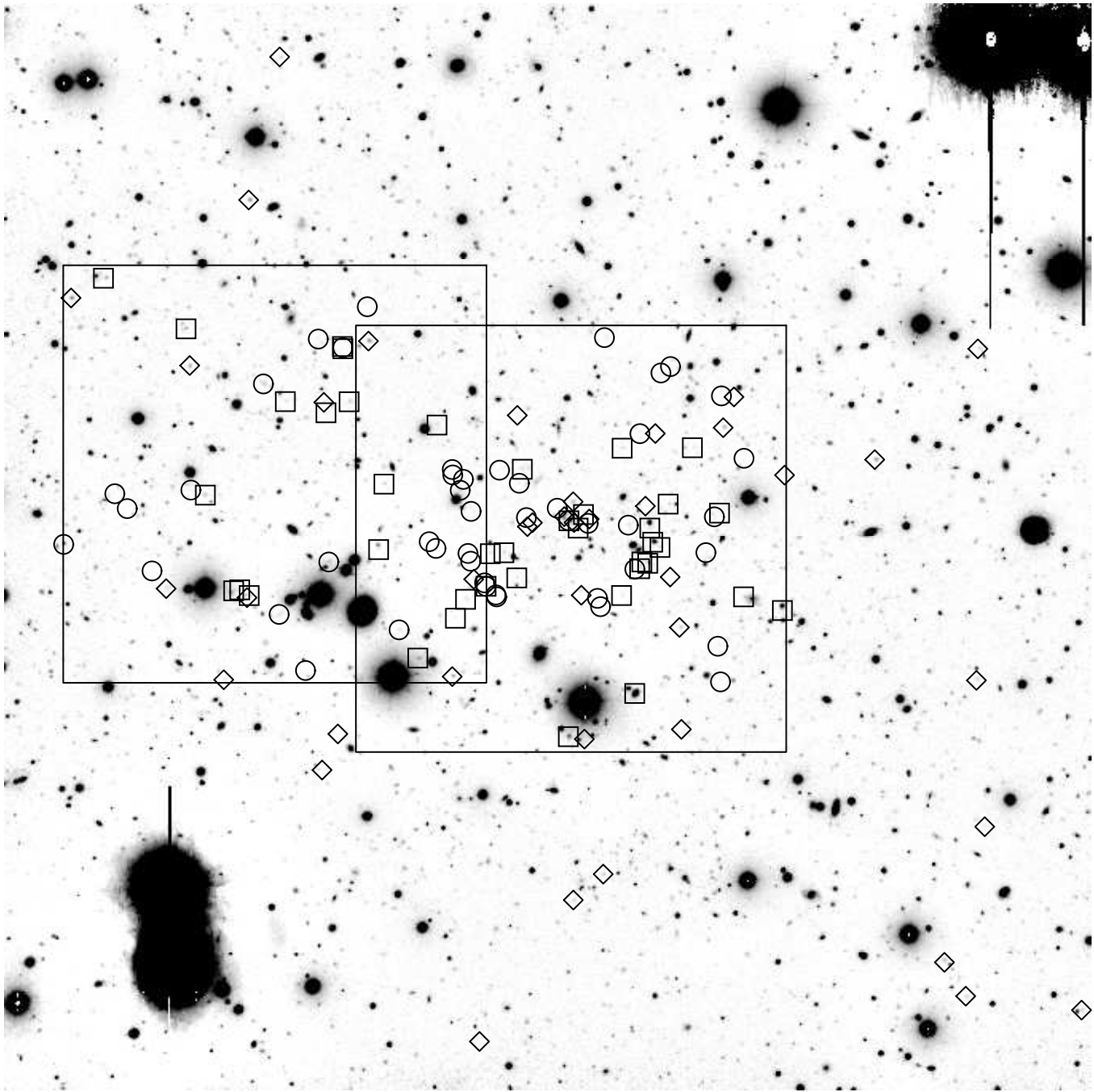}}
  \caption{A 6\farcm8$\times$6\farcm8 $I$ band image of the field of
  \rg. Candidate \lya\ emitters are indicated by diamonds, candidate
  \ha\ emitters by squares and EROs by circles. The two regions
  observed in \nbir\ and $K_s$ band are indicated by the boxes.}
  \label{I_emitters}
\end{figure*}

We have now identified candidate cluster members on the basis of three
different criteria and we are able to find out whether there is any
overlap between the three populations in the area covered by the
infrared imaging. The radio galaxy fulfills all requirements: it is an
extremely red object with both \lya\ and \ha\ emission. The extended
emission line halo also contains several objects which are found to be
either EROs, \ha\ or \lya\ emitters.  Apart from the radio galaxy,
there are two objects classified both as ERO and candidate \ha\
emitter. One of these is located in the central infrared field where a
$J$ magnitude is available and is placed within the starburst region
defined by \citet{poz00}. The \ha\ emitting EROs have $I - K$ magnitude
$\sim 4.4$. There are no candidate \lya\ emitters with $I - K$ colour
red enough to be classified as EROs. This is consistent with the idea
that the EROs are dusty starbursts \citep{dey99} for which we do not
expect \lya\ emission due to the strong extinction but at least some
\ha\ emission. There are several groups of EROs, \ha\ and \lya\
emitters close together, as can be seen in Fig.\
\ref{I_emitters}. None of the candidate \lya\ or \ha\ emitters have
sufficient infrared or optical narrow band excess emission to be
selected as a \ha\ or \lya\ candidate, respectively.  A more elaborate
discussion of \lya/\ha\ ratios of the candidate emitters is postponed
to Sect.\ \ref{lyaha_ratio}.

%In addition, there are three candidate \ha\ emitter at distances less
%then 3\arcsec\ from an ERO (1\arcsec\ is 8.93 kpc at $z = 2.16$). The
%chance on such a random superposition is small: from the surface
%density of EROs in the field, we expect $< 0.03$ EROs within a
%3\arcsec\ radius. These systems could therefore have a physical
%connection.  In Fig.\ \ref{I_emitters} we present an $I$ band image of
%the field of \rg\ indicating candidate \lya\ and \ha\ emitters and
%EROs.

\subsection{Comparison of the spatial distributions}
It seems (Figs.\ \ref{ero_circ}, \ref{ha_circ}, \ref{lya_circ}) that
the \ha\ candidates and EROs are more concentrated towards \rgs\ than
the \lya\ emitters. To determine whether these differences are
significant, we compare the concentration of galaxies towards the
radio galaxy by measuring the height and width of the density peak on
top of the background population.  We define the surface density of
the background population as the density at 60\arcsec\ (0.5 Mpc) $<$
R$_{1138} <$ 200\arcsec\ (1.8 Mpc), where R$_{1138}$ is the distance
from the radio galaxy. The heigth of the peak within certain values of
R$_{1138}$ is given in Table \ref{overdens} along with its
significance $\sigma$.  For this computation we have excluded the
radio galaxy halo objects.  It is clear that the density peaks of the
EROs and \ha\ candidates are more significant than of the \lya\
candidates for all R$_{1138} <$ 60\arcsec. The former peaks are also
more pronounced at short distances (R$_{1138} <$ 40\arcsec), i.e.\
their width is smaller.  We conclude that there is good evidence that
the EROs and candidate \ha\ emitters are more concentrated towards
\rg\ than the candidate \lya\ emitters.

\begin{table}
\begin{center}
\caption[]{Overdensities of the three samples} \label{overdens}
\begin{tabular}{l c c c c c c} \hline \hline
R$_{1138}$ &
\multicolumn{2}{r}{$<$ 40\arcsec} & 
\multicolumn{2}{r}{$<$ 60\arcsec} & 
\multicolumn{2}{r}{20\arcsec -- 60\arcsec} \\
\multicolumn{1}{c}{(1)} & (2) & (3) & (2) & (3) & (2) & (3) \\ \hline
%EROs & 5.3 & 2.9 & 3.5 & 3.2 & 3.3 & 3.3 \\
EROs & 4.7 & 2.5 & 3.2 & 2.9 & 3.3 & 3.3 \\ %nohalo
%\ha\ & 5.5 & 2.9 & 2.9 & 2.8 & 2.8 & 3.0 \\
\ha\ & 4.8 & 2.6 & 2.6 & 2.5 & 2.8 & 3.0 \\ %nohalo
%\lya & 6.1 & 2.4 & 4.4 & 2.6 & 2.7 & 1.6 \\
\lya & 2.6 & 1.0 & 3.1 & 1.5 & 3.1 & 1.7 \\ %nohalo
\hline \hline
\end{tabular}
\end{center}
%\begin{center}
\footnotesize \noindent Notes: (1) Sample, either EROs, candidate
\lya\ or \ha\ emitters (2) Density in terms of background density (3)
Significance ($\sigma$) of (2).
%\end{center}
\end{table}                     

\section{Properties of the candidates}\label{prop}

\subsection{\lya/\ha\ ratios of the candidate emitters}\label{lyaha_ratio}

\begin{figure}
  \resizebox{\hsize}{!}{\includegraphics{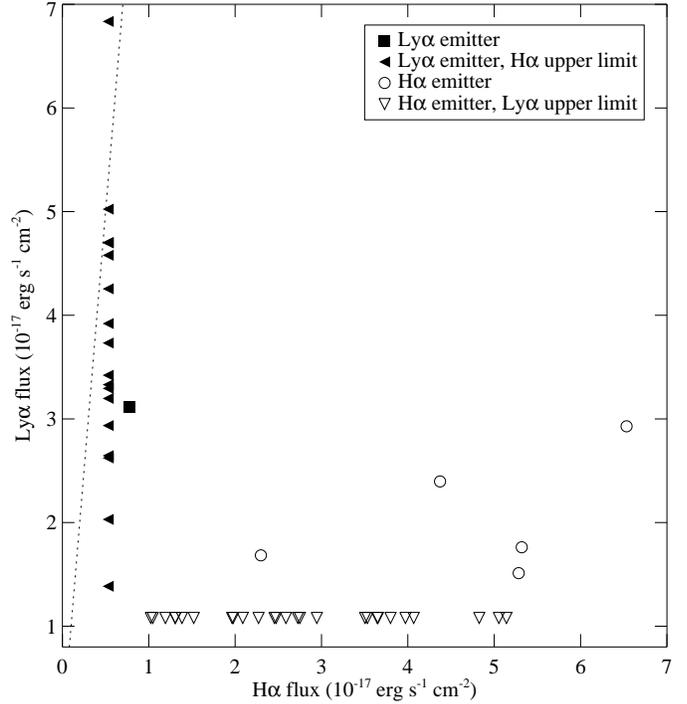}}
  \caption{\lya\ vs \ha\ flux for candidate \lya\ emitters (filled
  symbols) and \ha\ emitters (open symbols). Triangles indicate upper
  limits. The dotted line indicates a \lya/\ha\ ratio of 10,
  approximately the case B prediction.}
  \label{lyaha_plot}
\end{figure}

%\begin{figure}
%  \resizebox{\hsize}{!}{\includegraphics{ratio_vs_I.ps}}
%  \caption{\lya/\ha\ ratios vs $I$ band magnitude for candidate \lya\
%  emitters (filled symbols) and \ha\ emitters (open
%  symbols). Triangles indicate upper or lower limits. All emitters
%  plotted at \lya/\ha\ ratio of ten actually have values $>> 10$ and
%  all emitters plotted at $I = 26.8$ are not detected in the $I$ band
%  image.}
%  \label{ratio_vs_I}
%\end{figure}

%\begin{figure}
%  \resizebox{\hsize}{!}{\includegraphics{ratio_vs_K.ps}}
%  \caption{\lya/\ha\ ratios for both kinds of candidate emitters
%  against $K$ band magnitude.}
%  \label{ratio_vs_K}
%\end{figure}

There are no objects selected as both \lya\ and \ha\ candidates, but
for some candidates we have detected emission in the other line below
the selection criteria used in this work. The small overlap might seem
surprising at first since both lines are produced by the recombination
of neutral hydrogen. However, the strength of the \ha\ line is about a
factor of ten less than the \lya\ line in case B recombination
circumstances \citep{ost89}. Because the lowest \ha\ line flux detected
is $\sim 0.55 \times 10^{-17}$ \ecs, candidate \lya\ emitters should
have a \lya\ line flux in excess of $\sim 5.5 \times 10^{-17}$ \ecs\
to have a detectable \ha\ counterpart.

%one of the \lya\ emitter with a confirmed redshift (1240) is also in
%the \ha\ catalogue. This object is not selected as a \lya\ candidate
%due to its EW$_0$ being just below the selection criterion. It was
%included in the spectroscopy sample as a serendipitous object.

%If there is dust present in the emitter, this will absorb the
%resonantly scattered \lya\ emission very efficiently while the \ha\
%emission can escape unhindered in most cases. The presence of dust
%will therefore decrease the \lya/\ha\ ratio.

Of the 26 candidate \lya\ emitters with EW$_0 > 15$ \AA\ and $\Sigma >
3$ in the area covered by the infrared observations, nine have line
fluxes $> 5.0 \times 10^{-17}$ \ecs.  Six of the latter are part of
the extended \lya\ halo of \rgs.  We will discuss the other three
here in more detail.  For candidate 561 a line flux of $6.8 \times
10^{-17}$ \ecs\ was derived from the imaging observations, but
spectroscopic observations (described in paper II) indicate a flux of
$4.0 \times 10^{-17}$ \ecs. For this \lya\ flux level, we do not
expect to observe \ha\ emission. Candidate 778 has a \lya\ line flux
of $67.6 \times 10^{-17}$ \ecs\ (outside the range of fluxes displayed
in Fig.\ \ref{lyaha_plot}) and coincides with an object detected on
the \nbir\ image with a \ha\ flux of $8.8 \times 10^{-17}$ \ecs,
resulting in a \lya/\ha\ ratio of 7.7.  The object has an \ha\ EW$_0$
of 18 \AA\ and is therefore not included in the list of candidate \ha\
emitters. The emitter is further described in Sect.\ \ref{X-ray}.
Candidate 441 is part of a chain of emitters (confirmed in Paper II),
emitting both \lya\ and \ha, having a \lya/\ha\ ratio of 4.7. The
\lya\ emitter does not coincide with a candidate \ha\ emitter, because
there is no object detected at this exact location on the \nbir\ band,
but \ha\ candidate 145 is part of the same chain and only 1.3\arcsec\
away.

Only one candidate \lya\ emitter with \lya\ line flux below $5.5
\times 10^{-17}$ \ecs\ emits detectable \ha: candidate 675 has a \lya\
flux of $3.1 \times 10^{-17}$ \ecs\ and an \ha\ flux of $0.7 \times
10^{-17}$ \ecs, resulting in a ratio of 4.4 (see Fig.\
\ref{lyaha_plot}).  The object was not selected as an \ha\ candidate
emitter as its significance ($\Sigma$) is only 1.7.
The lower limits of the \lya/\ha\ ratio computed for the remaining
objects (see Fig.\ \ref{lyaha_plot}) are in the range 2--7.

%Candidate 561 has F$_{\rm Ly\alpha} = 6.8 \times 10^{-17}$ \ecs,
%halo. These objects have \lya/\ha\ ratios $>> 10$ which can be
%explained by the fact that \lya\ emission is scattered by neutral
%hydrogen to the outskirt of the hydrogen cloud encompassing AGN, while
%\ha\ is not (see the forthcoming paper on the emission line halo of
%\rgs, Kurk et al.\ in preparation). For the other candidate \lya\
%emitters, we have computed lower limits on the \lya/\ha\ ratio (see
%Fig.\ \ref{lyaha_plot}).  They are in the range 2--7, which implies
%that the \lya\ emitting galaxies are relatively dust free.
%Two (candidates 441 and 350) have \lya/\ha\ ratios of 4.65 and 6.14.

For 31 of the 40 candidate \ha\ emitters with EW$_0 > 25$ \AA\ and
$\Sigma > 2$ we detect no \lya\ emission. The upper limits for the
\lya/\ha\ ratio for these objects are in the range $0.03 - 1.06$.
Excluding three objects in the halo of \rgs, there are six \ha\
candidates for which \lya\ emission is detected. Their ratios are in
the range $0.10 - 0.73$ (see Fig.\ \ref{lyaha_plot}).

%The range of fluxes for the \ha\ candidates is $1.0 - 5 \times
%10^{-17}$ \ecs, except for 13 objects with fluxes in the range $10 -
%220 \times 10^{-17}$ \ecs. Two of these are part of the halo of \rgs\
%and one has a \lya\ flux of $19 \times 10^{-17}$ \ecs, implying a
%\lya/\ha\ ratio of 0.09. The others have no \lya\ emission detected in
%the optical images and have therefore upper limits on the \lya/\ha\
%ratio in the range $0.02 - 0.11$. Of the remaining 25 candidates, only
%one has detected \lya\ emission and this object is part of the \lya\
%halo of the radio galaxy. 

Note that the \ha\ and \lya\ emission for respectively the candidate
\lya\ and \ha\ emitters has been measured in the apertures defined on
respectively the \nbo\ and \nbir\ images. This can cause differences
in the \lya/\ha\ ratios for these objects within a factor two. This
discrepancy is, however, not large enough to explain the obvious
difference in ratios between the two types of candidates: candidate
\lya\ emitters have \lya/\ha\ ratios $> 2$, while candidate \ha\
emitters have ratio $< 1$. It is easily understood that objects
selected by \lya\ emission must have dust free sightlines and
therefore \lya/\ha\ ratios close to the case B value. There is
no {\bf such} selection bias towards low \lya/\ha\ ratios for objects
selected by \ha\ emission, but as star formation is generally
accompanied by dust production, it is not surprising that the
candidate \ha\ emitters have low \lya/\ha\ ratios.

The spatial positions of the candidates for which both \lya\ and
\ha\ has been detected are indicated on Fig.\ \ref{ha_pos} and Fig.\
\ref{lya_pos}.

%If both types of emitters were drawn from one underlying population of
%line emitting galaxies, we would expect the \lya/\ha\ ratios of the
%\ha\ emitters to cover the full range of possible values up to $\sim
%10$.  This bimodal distribution suggests that we observe two different
%populations of star forming galaxies.
% Fig.\ \ref{ratio_vs_I} shows the \lya/\ha\ ratio vs $I$ band magnitude 
% for candidate \lya\ and \ha\ emitters.

\subsection{Star formation rates}

\subsubsection{SFR estimators}
If clouds of neutral hydrogen in or near high redshift galaxies absorb
the integrated stellar light shortward of the Lyman limit and re-emit
this energy in nebular lines, such as \lya\ and \ha, they provide a
direct, sensitive probe of the young massive stellar population. Since
only stars with lifetimes shorter than 20 Myr contribute significantly
to the integrated ionizing flux of the galaxy, emission line flux is a
nearly instantaneous measure of the star formation rate (SFR). This
emission line flux can be computed by stellar population synthesis
models, but is very sensitive to the initial mass function (IMF)
assumed since almost exclusively stars with M $> 10$ M$_\odot$
contribute \citep{ken98}. Assuming a \citet{sal55} IMF with mass
limits 0.1 and 100 M$_\odot$ and solar metallicity, \citeauthor{ken98}
derives, using the evolutionary synthesis models of \citet{ken94}, the
following relation: \begin{equation} {\rm SFR~ (M_\odot yr^{-1}) = 7.9
\times 10^{-42} ~L_{H\alpha}~ (erg ~s^{-1})}. \label{sfrha}
\end{equation}

%Since the ratio of \lya\ to \ha\ emission of ionized
%gas is about ten in Case B recombination circumstances and only
%slightly dependent on the electron temperature, we can derive a
%similar relation for the \lya\ line: \begin{equation} {\rm SFR~
%(M_\odot yr^{-1}) = 7.9 \times 10^{-43} ~L_{Ly\alpha}~ (erg
%~s^{-1})}. \label{sfrlya} \end{equation}

Another way to measure the SFR is observing directly the rest frame
ultra violet light, which is dominated by young stars. The optimal
wavelength range is 1250 to 2500 \AA, longward of the \lya\ forest but
short enough that older stellar populations do not contribute
significantly. Using the same stellar population as described above,
for a model galaxy with continuous star formation during $\sim$ 100
Myr, \citet{ken98} finds that the luminosity in this wavelength region
scales directly with the SFR: \begin{equation} {\rm SFR~ (M_\odot
yr^{-1}) = 1.4 \times 10^{-28} ~L_\nu~ (erg ~s^{-1}
Hz^{-1})}. \label{sfruv} \end{equation} This conversion is, however,
dependent on the age of the stellar population and mode of star
formation. The SFR / L$_\nu$ ratio is higher\footnote{There is an
error in Sect.\ 2.2 of \citet{ken98}: the SFR/L$_\nu$ ratio will be
significantly \emph{higher} in younger populations (Kennicutt, private
communication).} in populations younger than 100 Myr, up to 57\% for a
9 Myr old population and lower in 100 Myr old populations with, for
example, an exponentially decreasing star formation rate.
\citet{gla99} extensively discuss the dependence of luminosity of the
\ha\ line, 1500\AA\ and 2800\AA\ continuum on age and metallicity of
the galaxy stellar population.

\subsubsection{SFRs of candidate \ha\ emitters}
We have computed the \sfrha\ from the \ha\ emission of the candidate
\ha\ emitters assuming they are at $z = 2.16$, correcting for a 25\%
contribution of the [\ion{N}{II}]6548+6584 \AA\ system \citep{ken83}.
For the IMF used to compute Eq.\ \ref{sfruv}, the UV spectrum happens
to be nearly flat in L$_\nu$ \citep{ken98}.  The central wavelength of
our $I$ band observations corresponds to 2430 \AA\ in the rest frame
of objects at $z = 2.16$ and is therefore suitable to estimate the
\sfruv\ of galaxies in the proto-cluster. Because we do not correct
for possible absorption by dust, both star formation estimators can be
considered lower limits to the intrinsic star formation in the
galaxies.

\begin{figure}
  \resizebox{\hsize}{!}{\includegraphics{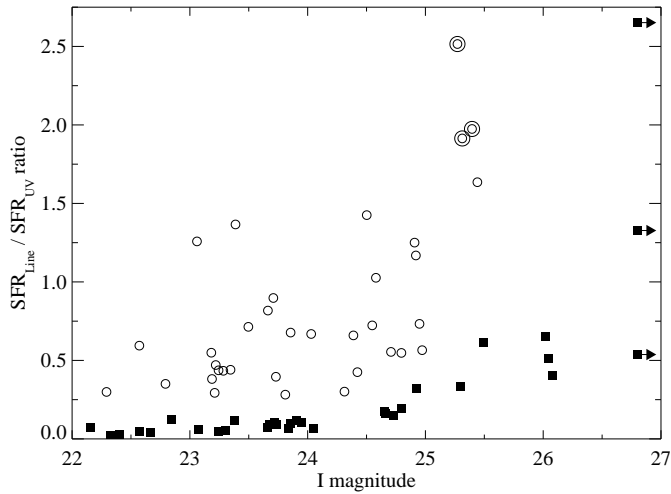}}
  \caption{The ratio of SFR derived from line emission to SFR derived
  from the rest frame UV continuum emission versus $I$ magnitude for
  candidate \ha\ emitters (open circles) and candidate \lya\ emitters
  (filled boxes).  Large circles indicate emitters coinciding with
  EROs.  The three \lya\ emitters plotted at $I = 26.8$ are not
  detected in the $I$ band. Components of \rgs, two known QSOs and one
  low redshift interloper are excluded from the plot.}
  \label{sfrlyaha_ratios}
\end{figure}

Excluding the radio galaxy components, a QSO (see Sect.\ \ref{X-ray})
and a low-redshift interloper (see Table \ref{Hacandidates}), the SFRs
derived from the \ha\ (UV) emission are in the range 2 -- 32 (3 -- 52)
M$_\odot$ yr$^{-1}$. The ratio of \sfrha / \sfruv\ is in the range 0.3
-- 2.5 with a mean of 0.8$\pm$0.1 and a dispersion of 0.5 (see Fig.\
\ref{sfrlyaha_ratios}).

Recently, \citet{bua02} have investigated the star formation rate
determined by the \ha\ line and the UV flux in a sample of nearby star
forming galaxies in clusters. They find a mean ratio \sfrha / \sfruv\
of 0.8 $\pm$ 0.4 and conclude that within the error bars the two SFR
estimators give consistent results. There is however a large scatter
in the sample, with two galaxies which exhibit an observed ratio of
$\sim$ 0.15. The mean ratio for an accompanying sample of 19 starburst
galaxies is $\sim$ 2, indicating that more dust is present in these
objects. These results were obtained with conversion factors for
luminosity to SFR significantly different from Kennicutt's values due
to a higher low mass cutoff of the IMF and the use of another
population synthesis program \citep[Starburst99,][]{lei99}, but the
ratio of SFR values is (coincidentally) exactly equal and their
results are therefore comparable with ours. Assuming different IMFs,
low and high mass cutoffs and periods since the burst for model
stellar populations, they obtain a theoretical range of ratios of 0.66
to 1.5 for dust free galaxies. Note that, although the methods give
identical values for the ratios, we would obtain values 50\% higher
using the equations in Sect.\ 4.1 of \citet{bua02} due to the $I$ band
sampling 2430 \AA\ in stead of 2000 \AA.

The ratios of \sfrha / \sfruv\ for our sample are consistent with the
values determined by \citet{bua02}. Ratios $> 1$ can be explained by
the presence of dust in the galaxies which extinguishes the UV
continuum more effectively than the \ha\ emission. The three
candidate \ha\ emitters which overlap with two EROs are probably the
most dusty objects in our sample (see Fig.\ \ref{sfrlyaha_ratios}).
The relatively strong UV emission of the \ha\ candidates with ratios
$< 1$ can indicate that the stellar population is older than 30 Myr,
when the \ha\ flux drops steeply, but UV flux presists up to 1 Gyr
\citep{gla99}.  Alternatively, these ratios could be caused by an
unconventional initial mass function or a contribution from direct or
scattered AGN light to the UV emission.  It is also possible that the
youngest stars which are responsible for the \ha\ emission are more
enshrouded in dust than the stars responsible for the bulk of the UV
emission \citep{moo00}.
%Some of the candidates might not be \ha\ emitters at $z \sim 2.2$. In
%that case, the ratio has an arbitrary value.

\subsubsection{SFRs of candidate \lya\ emitters}

%Since the ratio of \lya\ to \ha\ emission of ionized
%gas is about ten in Case B recombination circumstances and only
%slightly dependent on the electron temperature, we can derive a
%similar relation for the \lya\ line: \begin{equation} {\rm SFR~
%(M_\odot yr^{-1}) = 7.9 \times 10^{-43} ~L_{Ly\alpha}~ (erg
%~s^{-1})}. \label{sfrlya} \end{equation}

Assuming that the \lya\ emission we observe is produced in case B
recombination circumstances and not extincted by dust (as indicated by
the \lya/\ha\ ratios and lower limits computed in Sect.\
\ref{lyaha_ratio}), the strength of the \ha\ emission of the \lya\
emitters is one tenth that of the \lya\ emission. We use Eq.\
\ref{sfrha} to derive the SFR from line emission and Eq.\ \ref{sfruv}
from UV emission for the candidate \lya\ emitters.  Excluding the
radio galaxy, the QSO discovered in Paper II and the QSO described in
Sect.\ \ref{X-ray}, the range of \sfrlya\ is $0.4 - 4.3$
M$_\odot$yr$^{-1}$, while the range of SFR ratios is $0.02 - 2.65$
with a mean of 0.3$\pm$0.1 and a dispersion 0.5 (see Fig.\
\ref{sfrlyaha_ratios}). There are only two emitters with a ratio $>
1$. These are both not detected in the $I$ band.
%Two emitters, both not detected in $I$ band, have ratios $>$ 1. One of
%these is a spectroscopically confirmed proto-cluster members (Paper
%II, object 833). 
In general, the \sfrlya / \sfruv\ ratios of the \lya\ candidates are
much lower than those of the \ha\ candidates. This might be due to the
resonant nature of the \lya\ recombination line. Although we expect
the \lya\ candidates to contain almost no dust, \lya\ photons are
absorbed and reemitted many times by neutral hydrogen before they
escape the galaxies, increasing the chance to be absorbed by dust
enormously. In addition, the wavelength of \lya\ is more than 1000 \AA\
lower than the restframe wavelength of 2430 \AA\ which the $I$ band
samples and suffers therefore more from extinction. The derived values
of \sfrlya\ can therefore be considered as lower limits to the SFRs of
the candidates.

\subsubsection{SFRs derived from UV emission}

\begin{figure}
  \resizebox{\hsize}{!}{\includegraphics{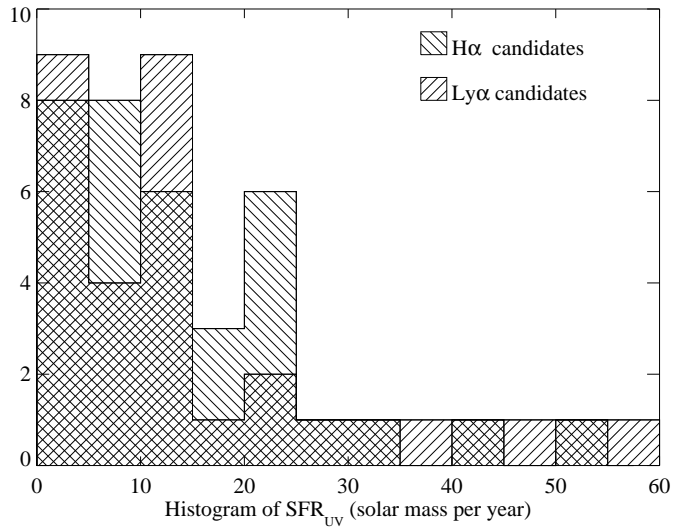}}
  \caption{A histogram of \sfruv\ for both \ha\ and \lya\ candidates.
  The horizontal axis shows the SFR in M$_\odot$yr$^{-1}$. The bin size
  is set to 5 M$_\odot$yr$^{-1}$. The radio galaxy, two QSOs and a low
  redshift interloper are excluded from this plot.}
  \label{sfruv_hist}
\end{figure}

To assess whether the samples of \lya\ and \ha\ candidates have
different SFRs, we consider the \sfruv\ for both kind of emitters
(Fig.\ \ref{sfruv_hist}), excluding components of the radio galaxy,
the two QSOs (see Sect.\ \ref{X-ray}) and one low redshift interloper
among the \ha\ sample.  To compute statistical properties of these
samples properly taking into account the non-detection, we have
employed survival statistics, using the methods presented in
\citet{fei85} as implemented in the ASURV (1.1) package \citep{iso90,
lav92}.  The range of inferred \sfruv\ of the \lya\ and \ha\ sample is
0.8 -- 59 \my\ with a mean of 16$\pm$3 \my\ and 2.9 -- 52 \my\ with a
mean of 14$\pm$2 \my, respectively.  The distribution of SFRs derived
from the $I$ band flux is therefore quite comparable for the two
populations.  In addition, a Peto-Prentice generalized Wilcoxon test
\citep{pre79} shows that there is a 62\% probability that the two
samples have the same underlying distribution.

%\medskip

Integrating the SFRs of the \lya\ and \ha\ emitters, we obtain values
of 507 and 501 \my, respectively. The comoving volumes in which we
have found these emitters are 9696 Mpc$^3$ ($2.110 < z < 2.163$, 43.6
arcmin$^2$) and 2076 Mpc$^3$ ($2.114 < z < 2.194$, 12.5 arcmin$^2$),
resulting in SFR densities (SFRDs) of 0.052 and 0.24
M$_\odot\,$yr$^{-1}\,$Mpc$^{-3}$, respectively.  Converted to an
Einstein-De Sitter cosmology with $H_0$ = 50 km s$^{-1}$ Mpc$^{-1}$
\citep[to compare with][]{mad96}, the SFRDs of \lya\ and \ha\ emitters
are 0.070 and 0.32 M$_\odot$yr$^{-1}$Mpc$^{-3}$, respectively, a
factor 4 to 18 more than the fiducial SFRD, as estimated by the mass
density of metals observed today divided by the present age of the
Universe, which is close to the lower limit of the SFRD at $z \sim
2.75$ (\citeauthor{mad96}). The observed SFRDs are about a factor ten
lower than the SFRD of the nearby Coma and Abell 1367 clusters, as
derived from the \ha\ luminosities of selected galaxies in these
clusters \citep{igl02}.  Note however, that the derived \ha\
luminosities of all our 35 \ha\ candidates are in excess of
$10^{41.5}$ erg s$^{-1}$, somewhat above the value of L$^*$ found for
Coma \citep[$10^{41.23}$ erg s$^{-1}$,][]{igl02}.  If the Coma cluster
were placed at $z = 2.16$, we would not have detected more than one of
its members by its \ha\ emission.

%Note, however that this value can be seen as a lower limit to the SFRD
%in the field of \rg, because we have not taken into account the
%contribution of emitters fainter than our selection criteria. The SFRD
%as derived from a proper integration of the luminosity function of
%emitters might yield a much higher value as indicated by a comparison
%with the luminosity function of \ha\ emitting galaxies in
%\citet{igl02}.
%The 35 \ha\ candidates all have \ha\ luminosities in excess of
%$10^{41.5}$ erg s$^{-1}$, indicating that the number of \ha\ bright
%galaxies per Mpc$^3$ near \rgs\ is about two orders of magnitude
%higher than observed in Coma or Abell 1367. If, however, not all
%considered candidates are emitters at $z \sim 2.2$, the resulting SFRD
%in the field of \rgs\ will be lower than presented here.

Another measure of the star formation activity is the ratio of SFR in
a galaxy to its mass. This ratio does not depend on the density of
galaxies in the (proto-) cluster.  We derive the stellar mass of the
candidate \ha\ emitters at $z = 2.16$ from their $K$ band magnitudes
as described in Sect.\ \ref{ssec:galmasses}. The ratio of star
formation rate to mass is therefore given by the EW$_0$: SFR / M
(yr$^{-1})$ = $7 \times 10^{-12}$ EW$_0$ (\AA). Although the Sloan
$r'$ filter used by \citet{igl02} samples a lower restframe wavelength
range (5440\AA\ -- 6755\AA) than our K$_s$ filter (6408\AA\ --
7263\AA), we assume here that the flat SED of galaxies above the
Balmer break justifies the use of the same formula for the Coma
cluster galaxies. There is no bias for EW$_0$ in the Coma cluster
sample, because all Coma galaxies with known velocities (up to $r'
\approx 16.5$) are selected. In our sample of \ha\ candidates, we have
only selected those with EW$_0 >$ 25\AA, so we have to exclude the
Coma cluster galaxies with EW$_0 <$ 25\AA. Twelve Coma galaxies remain
in the sample, which has a mean EW$_0$ of 76\AA, while the mean EW$_0$
of the non-AGN \ha\ candidates is 188\AA. This difference indicates
that the star formation rate per unit mass in the proto-cluster
galaxies is on average 2.5 times higher than in the Coma cluster
galaxies.

\subsection{Galaxy masses}\label{ssec:galmasses}

\begin{figure}
  \resizebox{\hsize}{!}{\includegraphics{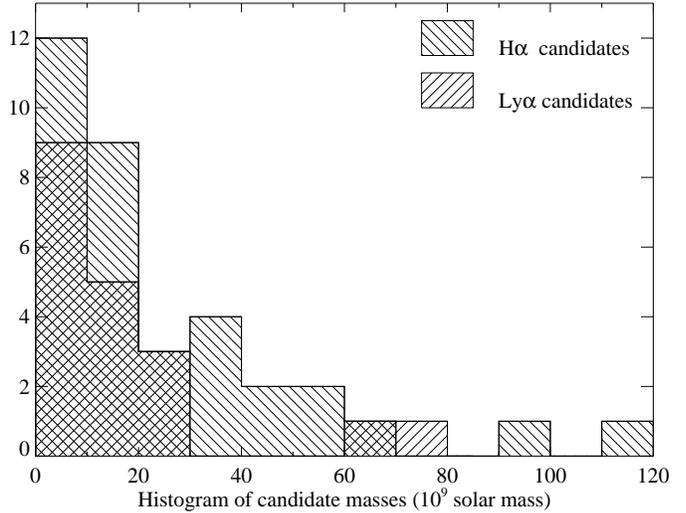}}
  \caption{A histogram of the $K_s$ derived masses for both \ha\ and
  \lya\ candidates. The horizontal axis shows the mass in $10^9$
  M$_\odot$.  The bin size is $10 \times 10^9$ M$_\odot$.}
  \label{mass_hist}
\end{figure}

Stellar masses for the galaxies are best estimated from infrared
magnitudes since at that wavelength range the influence of rare
short-lived high mass stars is minimal. The central wavelength of the
observed $K_s$ band magnitudes correspond to 6840 \AA\ in the rest
frame of the galaxies at $z = 2.16$. Since this wavelength is redward
of the 4000 \AA\ break, it is suitable to estimate galaxy masses
although maybe not optimal. We have computed an L/M ratio for galaxies
at $z = 2.16$ observed through this band using four models of GISSEL93
\citep{bru93} representing stellar populations with a constant SFR, an
exponential SFR and a 10 and 100 Myr burst. The IMF used in these
models is a \citet{sal55} law with lower mass cutoff at 0.1 M$_\odot$
and upper mass cutoff at 125 M$_\odot$. Assuming an age of 0.5 Gyr
($z_{\rm f} = 2.55$) for the stellar populations of the candidate
emitters, we obtain M/L ratios in the range of 2.5 -- 7.2
M$_\odot$ per L$_\odot$. We will use an M/L ratio of 4.8 M$_\odot$ /
L$_\odot$, which is the mean ratio of the four models. At 0.25 Gyr
and 0.75 Gyr the mean ratios are respectively 3.3 and 5.7
M$_\odot$ / L$_\odot$.  Note that the computed mass to light ratio of
$\sim 5$ in solar units is higher than commonly used ratios at high
redshift, because we do not use the bolometric luminosity of the
galaxies but only the luminosity in the restframe wavelength range
sampled by the K$_s$ band.

Within the area covered by the $K_s$ band images, there are 19 non-AGN
\lya\ candidates. Four of these (21\%) are not detected in $K_s$
band. Of the 35 non-AGN \ha\ candidates, also 4 (11\%) are not
detected in $K_s$ band. Taking into account the non-detections, the
mean flux densities expressed in $K$ magnitude of the \lya\ and \ha\
samples are respectively 21.2$\pm$0.3 and 20.7$\pm$0.2. We have
converted the $K$ magnitudes of the candidates to galaxy masses
assuming they are at $z = 2.16$, resulting in a range of masses of 3
-- 75 and 3 -- 113 $\times 10^9$ M$_\odot$ with means of 17$\pm$4 and
26$\pm$4 $\times 10^9$ M$_\odot$ respectively.  A histogram of masses
of the candidate emitters is shown in Fig.\ \ref{mass_hist}, which
shows that there are only two \lya\ emitters with M $>$ 30 $\times
10^9$ M$_\odot$, while there are ten \ha\ emitters in this range.  A
Peto-Prentice generalized Wilcoxon test \citep{pre79}, however, shows
that there is a 20\% probability that the two samples have the same
underlying distribution.  We conclude that there is tentative evidence
that the candidate \ha\ emitters are more massive than the candidate
\lya\ emitters.

The total mass of the non-AGN \lya\ and \ha\ emitters derived from the
detected $K$ band magnitudes is respectively 3.1 and 8.9 $\times
10^{11}$ M$_\odot$.

\subsection{X-ray properties}\label{X-ray}

\begin{figure}
  \resizebox{\hsize}{!}{\includegraphics{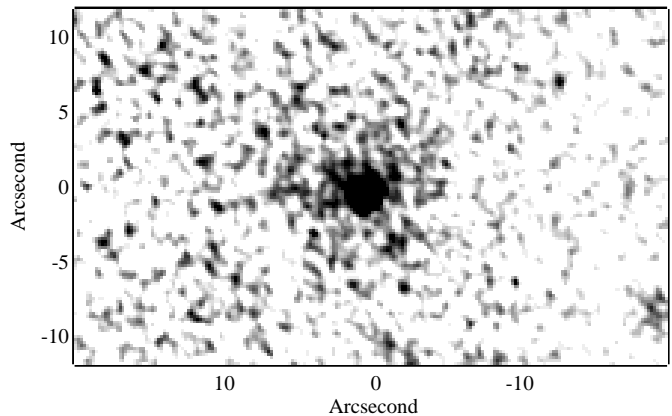}}
  \caption{Morphology of X-ray emitting candidate \lya\ emitter 778
  accompanied by faint extended \lya\ emission. The gray scale
  represents \lya\ line flux (narrow minus broad band), where white is
  the background level and black is 3 times the rms of the noise in
  the image. Coordinates of the central position are
  $\alpha,\delta_{\rm J2000}$ = ${\rm11^h41^m2\fs39}$,
  -26$^\circ$27\arcmin44\farcs6.}  \label{lya553}
\end{figure}

The properties of X-ray point sources in the field of \rg\ are
presented in a separate paper \citep{pen02}. Here, we only mention
that three X-ray point sources exhibit \lya\ emission and one \ha\
emission. One of these does not emit sufficient \lya\ emission to be
included in our catalogue (EW$_0$ = 3.2 \AA, $\Sigma = 3.6$). A second
(\lya\ candidate 968) is indeed spectroscopically confirmed to be a
QSO (see Paper II). A third (\lya\ candidate 778) has a very high
equivalent width (EW$_0 = 311$ \AA).  In addition, it displays a faint
\lya\ halo with a size of about 10\arcsec, visible in the convolved
image (Fig.\ \ref{lya553}). It was not included in the catalogue of
Paper I and has not been observed spectroscopically because it is not
covered by all jittered \nbo\ images. The presence of a halo, high EW,
high \lya\ line flux ($6.7 \times 10^{-16}$ \ecs) and X-ray emission
are consistent with an AGN nature of this object. The near infrared
spectrum of \ha\ candidate 215 emitting X-rays is presented in a
forthcoming paper, which confirms the AGN nature of this object.

\subsection{Conclusions on galaxy properties}

%Summary of SFR properties
%The SFRs derived from the line emission flux of the candidate \lya\
%emitters are lower than of the candidate \ha\ emitters. This is
%consistent with the resonant nature of the \lya\ line. A comparison of
%SFRs derived from the rest frame UV luminosity of both kinds of
%candidates shows that most are in the range of 2 - 60
%M$_\odot$yr$^{-1}$. The SFRD of the emitters in the field of \rgs\ is
%100 to 1000 times higher than estimates of the field SFRD at $z \sim 2$.

Using imaging observations in six bands, we have investigated several
galaxy populations possibly present in a proto-cluster at $z =
2.16$. From a selection on $I - K_s$ colour, we find several EROs. The
colours of these objects are not consistent with the colours computed
for evolved ellipticals at $z \sim 2$. However, the number density of
EROs is high and is increasing towards the radio galaxy, indicating
that some of the red objects must be associated with the radio galaxy
structure. The EROs could represent the progenitors of cluster
ellipticals with some star formation still going on. The presence of
excess emission consistent with \ha\ radiation at $z = 2.16$ from two
of the observed EROs supports this idea. A second sample is formed by
the objects selected on \nbir\ $- K_s$ colour. We believe that most of
these objects are \ha\ emitting galaxies at $z = 2.16$, associated
with the structure of galaxies around the radio galaxy. The increase
of the number density of these objects towards the radio galaxy again
supports this view. The high EWs of some candidate \ha\ emitters have
to be explained by an AGN contribution, but most emitters should be
powered by star formation with a rate of $10 - 100$ M$_\odot$
yr$^{-1}$. This is also true for the sample of candidate \lya\
emitters selected on the basis of \nbo\ $- B$ colour.

We have compared the properties of the \lya\ and \ha\ emitters.
The density of \ha\ emitters is higher close to \rg, their \lya/\ha\
ratios are lower than for the \lya\ emitters and their $K_s$ band
emission and implied masses are higher on average.  We propose
the following scenario to explain the observed differences. A larger
fraction of the \lya\ emitters is still being accumulated from the
environment as compared with the \ha\ emitters, while the more massive
\ha\ emitters have been able to retain more metal rich gas and dust
resulting in a lower \lya/\ha\ ratio.

%The \lya\ and \ha\ emitters have different properties and are
%therefore probably drawn from different populations of star forming
%galaxies. The density of \ha\ emitters is higher close to \rg, their
%\lya/\ha\ ratios are lower than for the \lya\ emitters and their $K_s$
%band emission and implied masses are higher on average. We propose the
%following interpretation for the difference in properties between the
%two types of candidates. The \lya\ emitters are very young galaxies,
%falling into the proto-cluster, while undergoing their first and
%virtually dust free starburst. The \ha\ emitters contain dust from
%previous starburst episodes and are therefore somewhat older and more
%settled in the proto-cluster potential than the \lya\ emitters.

\section{\rg: a rich distant cluster?}\label{clus}

In this section we investigate the cluster properties of the structure
associated with radio galaxy \rg.

\subsection{The overdensity of \lya\ emitters in redshift space}
\label{sec:overdens}
\begin{figure}[htbp]
  \resizebox{\hsize}{!}{\includegraphics{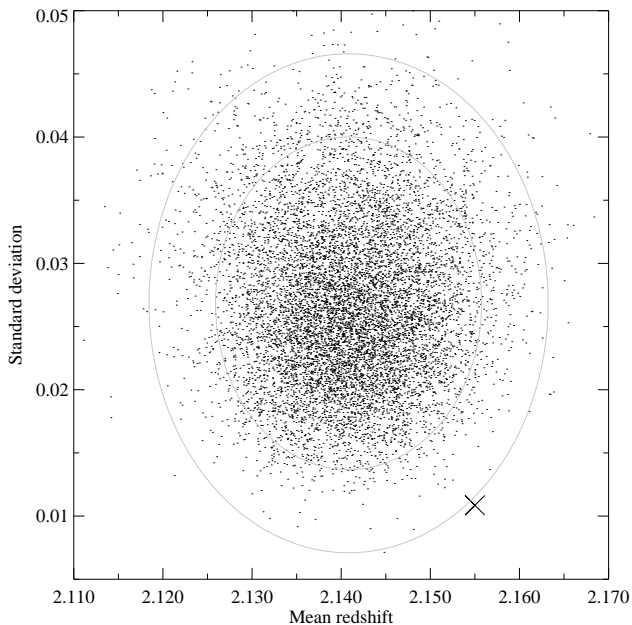}}
  \caption{The mean and standard deviations of the 10,000 simulated
  redshift distributions. The cross indicates the position of the
  measured redshift distribution of the confirmed \lya\ emitters and
  the contours indicate the 2 and 3$\sigma$ deviation from the
  combined mean and standard deviation from the simulated distributions.}
  \label{monte_zdist}
\end{figure}

In Paper II, we have presented the redshift distribution of the 15
confirmed \lya\ emitters. This distribution (see Fig.\ 3 in Paper II)
does not follow the sensitivity curve of the narrow band filter used
to select the candidate \lya\ emitters, but is narrower and centered
on the redshift of the radio galaxy, indicating that the \lya\
emitters have a physical connection to the central radio galaxy. To
quantify the significance of this deviation from randomness we have
carried out Monte Carlo simulations of the distribution of redshifts
observed through the narrow band filter. We have simulated 10,000
realizations of the redshift distribution of 14 \lya\ emitters. The
redshift of each simulated emitter was chosen according to the
probability function derived from the sensitivity of the \nbo\ filter
and the VLT overall efficiency.  The mean and standard deviation of
each realization is plotted in Fig.\ \ref{monte_zdist}. The mean ($z =
2.155$) and standard deviation ($\delta z = 0.0109$) of the redshift
distribution of the confirmed \lya\ emitters is indicated by a cross
on this figure, which diverge respectively 2.0$\sigma$ and 2.4$\sigma$
from the simulated values\footnote{Note that the narrow band filter
was designed for the cone angle of FORS1 and its transmission was
measured within the instrument. The measured deviation of the mean
redshift can therefore not be due to the dependence of interference
filter specifications on the incident beam.}.  The measured
distribution deviates therefore 3.1$\sigma$ from a random
distribution, meaning that the probability that the redshifts of the
confirmed \lya\ emitters are drawn from a such a distribution is less
than 0.2\%. This shows that the ensemble of \lya\ emitters must have a
physical connection with the central radio galaxy.

\subsection{Mass and overdensity of the proto-cluster}\label{clustermass}
In Paper II the mass of the proto-cluster and volume density of
the \lya\ emitters was shortly discussed. A more thorough discussion
follows here.

We can consider two limiting cases of the dynamical state of the
structure of galaxies: it is detached from the Hubble flow and
completely virialized or it is no dynamical entity at all and the
galaxies move according to the expansion of the Universe. In the first
case, the velocities of the galaxies are representative of the
potential well of the cluster, while in the latter case the redshifts
are directly related to distances.

\medskip

In the first case, we can compute a virial mass as an estimate of the
cluster mass. The group of galaxies at $z \sim 2.16$ consists of 15
members, including \rg\ but excluding the QSO at $z = 2.183$,
which is a $3\sigma$ outlier. The redshift distribution of the members
is bimodal around the radio galaxy (see Paper II). The
velocities $v_i$ of the galaxies relative to the mean velocity of the
group are computed in the following way: \begin{equation} v_i = c~(z_i
- \bar{z}) ~/~ (1 + \bar{z}). \end{equation} The velocity dispersion
of the two groups of 7 members each (the radio galaxy is excluded in
this case) is 385 \kms\ and 204 \kms, as computed by the Gapper sigma
method \citep{bee90}, which is considerably smaller (23\%) than
computed by the standard method (square root of variance). The
velocity dispersion of the entire sample is 901 \kms, as computed by
the Gapper sigma method, which is 10\% smaller than by the standard
method.

The virial radius $R_{\rm v}$ and mass M$_{\rm v}$ of N cluster
members at positions $r_i$ with velocity dispersion $\sigma_v$ is
computed as follows:
\begin{equation} R_{\rm v} = \pi~ \frac{1}{2}~ \frac{\rm N~(N-1)}{2} 
\biggl( \sum_{i<j}\frac{1}{|r_i - r_j|} \biggr)^{-1}, \end{equation}
\begin{equation} {\rm M_v} = \frac{6}{\rm G}~ \sigma_v^2 ~R_{\rm v}.
\end{equation} 
\citet{sma98} show that the virial mass estimator works reasonable
well even for a bound system with substantial substructure. If the
system is only marginally bound, the estimate is too high but not more
than a factor of two. The largest uncertainties in our estimate are
the unknown dynamical state (i.e.\ one group or two groups, bound or
unbound) and the small number of available redshifts.

The virial radii of the two groups are 0.8 and 1.1 Mpc, about two
thirds of the maximum observable radius of 1.5 Mpc which was fixed by
the size of the CCD. The virial masses of the groups are therefore 1.7
and 0.6 $\times 10^{14}$ M$_\odot$. The mass of the two groups
together (2.3 $\times 10^{14}$ M$_\odot$) is much smaller than the
virial mass computed for the group of 14 galaxies as a whole (14
$\times 10^{14}$ M$_\odot$).

\medskip

In the second case, we assume that the emitters just detached from the
Hubble flow and are now collapsing to a common center of gravity. In
this case the redshifts of the emitters can be used as distance
indicators with a small correction for their motion towards each
other.  We can compute the comoving space density of \lya\ emitters,
its associated galaxy over-density and from this, derive a mass
over-density.  To compute the space density, we only consider the 14
confirmed emitters of Paper II in the redshift range $2.139 \le z \le
2.170$. The (uncorrected) comoving volume corresponding to this range
is 5611 Mpc$^3$. The comoving volume density of the confirmed emitters
is therefore 0.0025 Mpc$^{-3}$, a factor 1.4 smaller than the density
found in for the LBG peak at $z = 3.09$ (S00).  Because not
all candidates were observed and some were too faint to produce a
detectable line in their spectra, the resulting over-density of
4.4$\pm$1.2 with respect to the field population of \lya\ emitters is
a lower limit.

The total mass of the structure can be estimated by following \citet{ste98}
\begin{equation} {\rm M} = \overline{\rho}~ V~ (1 + \delta_{\rm m}) = 6.6
\times 10^{15} {\rm M}_\odot~ (1 + \delta_{\rm m}) \end{equation} with
$\overline{\rho}$ the mean density of the universe and $\delta_{\rm
m}$ the mass over-density within the volume containing the confirmed
emitters.  The mass density is related to the galaxy over-density
$\delta_{\rm gal}$ through
\begin{equation} 1 + b\delta_{\rm m} = C~ (1 + \delta_{\rm gal}),~~ C =
1 + f - f(1 + \delta_{\rm m})^{(1/3)} \end{equation} where $b$ is the
bias parameter, $C$ takes into account the redshift space distortions
caused by peculiar velocities and $f$ is the velocity factor
\citep{pee93}. This factor depends on the redshift and its value is
0.96 for $z = 2.16$. From the statistics of peaks in the redshift
distribution of LBGs, \citet{ste98} argue that $b \ge 4$. Taking $b$
in the range 3--5, $\delta_{\rm m}$ is estimated to be 0.6--1.2. This
implies a mass of our structure of $(1.1-1.5) \times 10^{16}$
M$_\odot$. 

%The mass of about $10^{16}$ M$_\odot$ found for the structure
%is in the range of the super-cluster structures. The mass of the Corona
%Borealis super-cluster at $z = 0.07$ including seven known Abell
%clusters, for example, is estimated by \citet{sma98} to be between 3
%and 8 $\times 10^{16}$ M$_\odot$.

%Note that the above computation assumes that all mass within
%the volume defined by the redshift distribution will contract into a
%cluster with a radius of about 1 Mpc, which is a contraction by a
%factor 15 in all three dimensions.

%The two methods to estimate the cluster mass produce results which
%diverge by a factor of $\sim 50$. 

The age of the universe at $z = 2.155$ is 3.7 Gyr, while the crossing
time for each group is about 3 Gyr. It is therefore impossible for the
system around \rgs\ to be in a relaxed state since it takes a few
crossings of the cluster members to virialize.  Hubble flow
expansion of the galaxies could lead to an artificially enlarged
\emph{velocity dispersion}. The \emph{virial} mass of $1.4 \times
10^{15}$ M$_\odot$ must therefore be considered an upper limit to the
mass of the bound (part of the) system.  Similarly, part of the mass
computed with the second method is not bound to the system.  Following
the same procedure for the used volume without an overdensity results
in a mass of $6.5 \times 10^{15}$ M$_\odot$.  This mass is mostly
intergalactic gas which will disappear out of the cluster with the
Hubble flow.  A more conservative estimate of the bound mass at $z =
2.16$ is therefore $4 - 8 \times 10^{15}$ M$_\odot$, although the true
bound mass of the system is probably much lower, which does not
exclude the possibility that more mass wil be bound at a later time.

%It is however also improbable that the cluster members have not yet
%detached from the Hubble flow, given the fact that a massive galaxy
%has already formed in their midst.

%The actual mass of the proto-cluster must lie in the broad range given
%by the two estimates. Note that the minimum mass of the proto-cluster
%is two orders of magnitude higher than the total stellar mass derived
%from the $K_s$ band magnitudes of the candidate line emitters ($1.2
%\times 10^{12}$ \my).

\subsection{Conclusions on cluster properties}

Using the lists of clusters members selected in the first part of this
paper, we have investigated the cluster properties of the galaxies in
the field of \rg. About 5\% of the galaxies in the local universe are
gathered in groups or clusters whose space density is larger than one
galaxy per cubic megaparsec, about two orders of magnitude greater
than the average density \citep{dre84}. Dense and populous clusters
\citep[as defined and cataloged by][]{abe58} contain on the order of
100 galaxies within two orders of magnitudes of the third brightest
member. The cluster members are gravitationally bound and to a large
extent in dynamical equilibrium (i.e.\ virialized). Because the
universe is about 4 Gyr old at $z = 2.2$, we do not expect large
virialized systems by that time and we will therefore call the
over-density of objects around \rg\ a proto-cluster which will develop
into a cluster. The evidence for an over-density of galaxies associated
with the radio galaxy comes from the high density of $K_s$ band
galaxies and ERO counts as compared with field counts and the
indication that most of this over-density is located in a 40\arcsec\
region around the radio galaxy. The density of $K$ band sources near
the radio galaxy corresponds to the density of current day's clusters
richer than class 0 or 1. Furthermore, the density of candidate \ha\
emitters is similar or higher than the density found around other
known high redshift (active) galaxies and is clearly highest in the
40\arcsec\ region around \rg.  Finally, the comoving volume density of
candidate \lya\ emitters in the field of the radio galaxy is higher
than in a blank field, if compared to the number of \lya\ emitters in
the field at $z = 2.4$ and 3.4 \citep{sti01,cow98} and in the
over-density of LBGs at $z = 3.09$ found by \citet{ste00}. From the
imaging alone, we conclude that \rg\ is located in an over-density of
galaxies, while the spectroscopical observations provide additional
confirmation. The 14 \lya\ emitters with redshifts confirmed by
spectroscopy have a redshift distribution which shows that they are
clearly associated with the radio galaxy. The probability that we have
observed a rare realization of randomly distributed emitters is less
than 0.4\%. The comoving volume density of the confirmed emitters is
already higher than in the field and is only a subset of the emitters
in the volume. We therefore conclude that \rg\ is located in a density
peak which will most probably evolve into a cluster of galaxies. 

%The mass of this cluster is estimated to be in the broad range of $2
%\times 10^{14}$ to $10^{16}$ M$_\odot$, mainly depending on the
%assumed dynamical state of the observed structure.

%The mass of this cluster can be estimated by different methods. The
%redshift and spatial distribution of the emitters yield a virial mass
%of $\sim 2 \times 10^{14}$ M$_\odot$ for the proto-cluster
%system. This estimate is a lower limit since the cluster members have
%not had enough time to virialize. More probably they are still falling
%in to the barycenter, causing the velocity dispersion to look smaller
%than it actually is. Another way to estimate the mass of the system is
%to measure the over-density in a particular volume and assume that all
%mass is gravitationally bound and will therefore contract to form a
%cluster of galaxies. This estimate yields a value of $\sim 10^{16}$
%M$_\odot$, a mass comparable with the mass of current day's
%super-clusters. The actual mass of the structure around \rg\
%lies probably somewhere in between these estimates.

An indication that the proto-cluster is not virialized comes from the
lack of extended X-ray emission perpendicular to the radio sources
axis \citep{car02}. These observations imply an upper limit to the
2--10 keV luminosity of an extended relaxed cluster atmosphere of $1.5
\times 10^{44}$ erg s$^{-1}$, less than 40\% of the X-ray luminosity
of the Cygnus A cluster.  Given the unrelaxed state of the system,
it is not possible to make an estimate of the bound mass at $z =
2.16$, but an upper limit of $\sim 10^{15}$ M$_\odot$ is computed for
a virialized system with the properties of the \rg\ proto-cluster.

\begin{figure}
  \resizebox{\hsize}{!}{\includegraphics{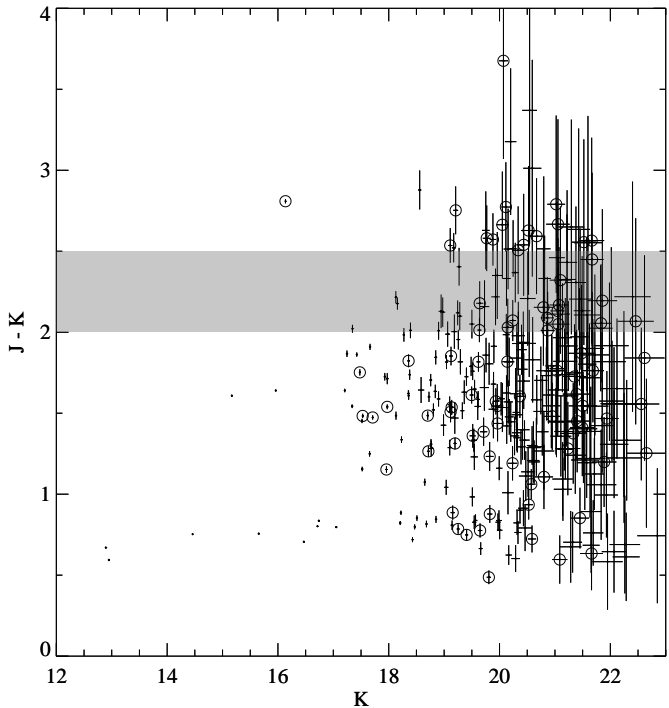}}
  \caption{Colour-magnitude plot of $J - K$ vs $K$ for the 306 bona
  fide sources detected on the $K_s$ and $J$ band images. The sources
  within 40\arcsec\ of the radio galaxy are indicated by circles and
  the expected locus of the sequence of passive cluster ellipticals at
  $z = 2.2$ is indicated by the grey bar. The $J - K$ locus is
  computed from the model defined in Sect.\ \ref{sec:eros_proto} (see
  also Fig.\ \ref{jkik_ero}).}
  \label{cm}
\end{figure}

Finally, we consider the galaxy population in the proto-cluster.  As
discussed in Sect.\ \ref{ero_sec}, there is a population of EROs
present in the cluster, most plausibly identified with dusty star
forming galaxies.  Contrary to what is observed in low redshift
clusters, we do not observe a sequence due to passive ellipticals in
the $J - K$ vs $K$ colour-magnitude diagram, presented in Fig.\
\ref{cm}. Both results are consistent with the idea that cluster
galaxies have evolved and were more luminous and bluer at high
redshift \citep[e.g.\ ][]{dok98}.

%The number of galaxies with absolute UV and $r$ band magnitudes $<
%-22$ and -21 is much larger than expected for a cluster like Coma at
%$z = 2.2$, as derived from the $R$ and $K$ band magnitudes of the
%candidate line emitters.

%Since we only observe part of the proto-cluster population with a
%strong bias towards star forming galaxies, we cannot make a direct
%comparison of the SFR in clusters at lower redshift. 

%The range of bona fide SFRs (1 -- 60 M$_\odot$ yr$^{-1}$) of the
%candidate \ha\ and \lya\ emitters shows that the proto-cluster
%contains a significant number of strongly star forming galaxies. In
%Sect.~\ref{prop} we have suggested that the candidate \lya\ emitters
%comprise a younger population of galaxies than the candidate \ha\
%emitters. The spatial distribution of the \lya\ emitters which is
%quite uniform over the field seems to support this idea. We surmise
%that the \lya\ emitters are still falling in, while the \ha\ emitters
%which are concentrated towards the radio galaxy have settled a bit
%more. This picture is similar to more evolved low redshift clusters in
%which more blue galaxies are found in the outskirts of the clusters
%than in their central regions.

\section{Summary}\label{sum}

We have presented new near infrared and optical imaging of the field
of radio galaxy \rg\ at $z = 2.16$. From these data and
published optical imaging and spectroscopy, we find several galaxy
populations associated with the radio galaxy.

$\bullet$ In the 12.5 square arcminute field covered by the $K_s$ band
images, we detect 550 objects in $K_s$. The number density of $K$ band
sources is a about 1.5 times higher than blank field counts and
increases towards the radio galaxy.

$\bullet$ In the same area, we find 44 objects with $I - K > 4.3$. The
surface density of EROs is about a factor two higher than the mean
density found in blank fields. Because the density of EROs increases
towards the radio galaxy, we conclude that some of the EROs are
galaxies at $z \sim 2.2$ and associated with the radio galaxy. Most of
these are star forming galaxies reddened by dust, although up to five
may be passive ellipticals.

$\bullet$ Also in the same area, we find 40 candidate \ha\ emitters.
The density of the brightest emitters is comparable to the density in
other AGN fields and, again, increases towards the radio galaxy. The
upper limit to the \lya/\ha\ ratio of these sources is 1.1, but for
most candidates the ratio is much smaller, meaning that these objects
contain some dust or neutral hydrogen. Two \ha\ candidates are also
identified as EROs. These must be dusty starbursts at $z \sim 2.2$.

$\bullet$ In the effective 43.6 square arcminute area of the optical
images, we find 40 candidate \lya\ emitters. The surface density of
these candidates does not show an increase towards the radio
galaxy. The \lya/\ha\ ratio of the 26 \lya\ candidates in the area
covered by infrared imaging is in the range $2 - 7$. The sight lines
through which we observe the \lya\ radiation are therefore relatively
dust free.

$\bullet$ The SFRs derived from rest frame UV flux of the sample of
\ha\ and \lya\ candidates are comparable, but the galaxy masses
derived from the flux around $\sim 7000$\AA\ are on average larger for
the \ha\ candidates than for the \lya\ candidates. From the spatial
distribution of the two samples, we deduce that the \ha\ emitters are
more settled into the potential well of the proto-cluster than the
\lya\ emitters.

%These properties may imply that the \ha\ emitters are older galaxies
%than the \lya\ emitters, although still young compared with present
%day cluster ellipticals.

%$\bullet$ A comparison of the proto-cluster at $z = 2.2$ with the Coma
%cluster luminosity function indicates that cluster populations evolve:
%the proto-cluster members at $z \sim 2.2$ are more luminous than their
%low redshift counterparts.

%$\bullet$ The bimodal redshift distribution of the confirmed sample of
%\lya\ emitters is consistent with the view that powerful radio sources
%at $z \ge 1$ are triggered by the merging of two cluster structures.

$\bullet$ From the spatial distribution, number density and redshift
distribution of objects detected in the field of \rg, we conclude that
the radio galaxy is located in a density peak which will evolve into a
cluster of galaxies at the present day.  The bound mass at $z =
2.16$ is $10^{15}$ M$_\odot$ at most.

%with a mass between $2 \times 10^{14}$ and $1 \times 10^{16}$
%M$_\odot$.
%The sheer existence of a structure with such a mass at $z \sim 2.2$ 
%rules out a cosmology with $\Omega_{\rm M} = 1$.

\smallskip
The structure around \rg\ is one of the few proto-clusters known at $z
> 2$ and deserves further study. X-ray observations of this field have
been carried out by \citet{car02} and an analysis of the X-ray point
sources and AGN content of the proto-cluster is presented in
\citet{pen02}.  Infrared spectroscopy of a small sample of the
candidate \ha\ emitters has been carried out and will be reported in
Kurk et al.\ (in preparation).

\begin{acknowledgements}
We acknowledge productive discussions with B.\ Venemans, M.\ Jarvis
and J.\ Fynbo. J.\ Fynbo kindly provided the algorithm for the Monte
Carlo simulation of \lya\ emitter redshift distributions (Sect.\
\ref{sec:overdens}). We thank the referee for its comments which have
improved the paper. This research has made use of the NASA/IPAC
Extragalactic Database (NED) which is operated by the Jet Propulsion
Laboratory, California Institute of Technology, under contract with
the National Aeronautics and Space Administration. We have also made
use of NASA's Astrophysics Data System Bibliographic Services.
\end{acknowledgements}

\bibliographystyle{aa}
\bibliography{ms3501.bib}

\appendix
\section{Source lists}
\begin{table}[htbp]
\caption{Candidate \ha\ emitters with EW$_0 > 25$ \AA\ and $\Sigma > 2$}
\begin{center} \label{Hacandidates}
\begin{tabular}{rrrrrr} \hline \hline
ID &  \multicolumn{1}{c}{Coordinates} & \multicolumn{1}{c}{$K$}
& EW$_0$ & \multicolumn{1}{c}{$\Sigma$} & F$_{\rm H\alpha}$\\
(1) & \multicolumn{1}{c}{(2)} & \multicolumn{1}{c}{(3)} & 
(4) & \multicolumn{1}{c}{(5)} & (6)\\ \hline
       5& 11 40 48.4  -26 30 30.6& 20.3&  89.6&   4.4&   6.5\\
  $^a$29& 11 40 46.5  -26 30 14.1& 17.7&  25.4&  15.9&  21.3\\
      79& 11 40 52.6  -26 30 01.0& 20.9& 120.2&   5.2&   4.8\\
     117& 11 40 51.6  -26 29 45.9& 21.3&  73.5&   3.0&   2.1\\
     131& 11 40 51.3  -26 29 38.7& 19.3&  56.5&   9.5&  10.9\\
     132& 11 40 42.4  -26 29 42.6& 19.9&  45.8&   4.2&   5.1\\
     144& 11 40 43.5  -26 29 37.4& 21.5& 246.5&   6.2&   5.1\\
     145& 11 40 57.4  -26 29 37.5& 20.3&  60.9&   3.4&   4.4\\
     152& 11 40 46.9  -26 29 37.1& 23.0& 577.2&   4.1&   2.3\\
     154& 11 40 50.7  -26 29 33.7& 20.3&  47.0&   3.4&   3.5\\
     155& 11 40 57.8  -26 29 35.7& 20.1&  30.1&   2.1&   2.7\\
     158& 11 40 57.6  -26 29 35.4& 21.1&  56.9&   2.3&   2.0\\
     167& 11 40 49.8  -26 29 30.5& 23.0& 278.4&   2.2&   1.4\\
     176& 11 40 46.4  -26 29 26.9& 20.1&  43.4&   3.0&   4.0\\
     183& 11 40 46.2  -26 29 24.9& 20.4& 165.0&  12.9&  10.0\\
     192& 11 40 46.3  -26 29 24.4& 19.7&  26.9&   3.2&   3.6\\
 $^b$199& 11 40 48.4  -26 29 08.9& 16.1&  63.9& 
            \makebox[1ex][r]{110.2}& \makebox[1ex][r]{221.9}\\
     207& 11 40 50.2  -26 29 21.0& 21.0&  95.1&   3.0&   3.5\\
     210& 11 40 50.6  -26 29 21.4& 22.6& 139.5&   2.4&   1.2\\
     211& 11 40 53.7  -26 29 20.0& 22.0&  64.1&   2.0&   1.0\\
     212& 11 40 45.8  -26 29 18.9& 21.7& 155.2&   3.6&   2.9\\
 $^c$215& 11 40 46.0  -26 29 16.9& 18.7&  99.9&  36.3&  31.5\\
     229& 11 40 46.1  -26 29 11.5& 19.1&  61.0&  11.2&  13.5\\
 $^b$269& 11 40 48.1  -26 29 11.6& 20.5& 167.3&   6.5&   9.5\\
 $^b$272& 11 40 48.0  -26 29 06.5& 19.9&  38.1&   4.3&   4.3\\
     279& 11 40 44.1  -26 29 05.8& 22.6& 263.7&   2.8&   2.0\\
     284& 11 40 45.6  -26 29 02.4& 22.4& 126.4&   2.4&   1.3\\
     293& 11 40 58.6  -26 28 59.6& 21.2& 115.8&   2.9&   3.7\\
     310& 11 40 53.6  -26 28 55.2& 21.8&  84.5&   2.0&   1.5\\
     318& 11 40 49.7  -26 28 49.4& 23.0& 257.8&   2.1&   1.3\\
     329& 11 40 46.9  -26 28 41.4& 23.0&
                       \makebox[1ex][r]{2222.2}&  4.8&   4.1\\
     356& 11 40 44.9  -26 28 41.1& 22.2&  81.0&   2.0&   1.0\\
     375& 11 40 52.1  -26 28 32.8& 22.3& 233.7&   2.7&   2.3\\
     380& 11 40 55.2  -26 28 28.3& 22.9& 547.1&   3.0&   2.5\\
     388& 11 40 56.4  -26 28 24.1& 21.3& 100.0&   3.8&   2.7\\
     394& 11 40 54.6  -26 28 24.1& 19.8&  32.1&   3.3&   3.8\\
     431& 11 40 54.8  -26 28 04.0& 20.9&  60.6&   2.3&   2.5\\
     437& 11 40 54.8  -26 28 03.2& 21.1&  71.2&   2.4&   2.6\\
     457& 11 40 59.2  -26 27 56.6& 20.9& 137.4&   8.4&   5.3\\
     477& 11 41 01.5  -26 27 37.6& 21.0& 154.1&   3.5&   5.3\\
\hline \hline
\end{tabular}
\end{center}
%\begin{center}
\footnotesize \noindent Notes: (1) Catalog number (2) Right ascension
and declination in J2000 coordinates (3) $K$ band magnitude (4) Rest
frame equivalent width (5) Significance of excess flux (6) Continuum
subtracted narrow band flux in $10^{-17}$ erg cm $^{-2}$
s$^{-1}$. Notes to individual objects: $^a$ Magnitude and morphology
indicate that this object is a low redshift interloper, $^b$ Inside
\lya\ halo of \rg, $^c$ QSO (see Sect.\ \ref{X-ray}).
%\end{center}
\end{table}

\begin{table}[htbp]
\caption{Candidate \lya\ emitters with EW$_0 > 15$ \AA\ and $\Sigma > 3$}
\begin{center}
\begin{tabular}{rrrrrr} \hline \hline \label{Lyacandidates}
ID &  \multicolumn{1}{c}{Coordinates} & \multicolumn{1}{c}{$B$}
& EW$_0$ & \multicolumn{1}{c}{$\Sigma$} & F$_{\rm Ly\alpha}$\\
(1) & \multicolumn{1}{c}{(2)} &  
\multicolumn{1}{c}{(3)} & (4) & (5) & (6)\\ \hline
       28& 11:40:50.8  -26:32:26.0& 26.1&  97.8&   4.3&   5.3\\
       46& 11:40:33.9  -26:32:13.4& 25.5&  31.6&   3.7&   3.4\\
       54& 11:40:37.1  -26:32:08.3& 24.5&  24.2&   5.6&   7.1\\
       73& 11:40:37.8  -26:31:55.5& 25.7&  34.5&   3.6&   3.1\\
      127& 11:40:48.2  -26:31:32.4& 25.5&  18.9&   3.3&   2.2\\
      146& 11:40:47.4  -26:31:22.6& 25.6&  26.0&   3.5&   2.6\\
      184& 11:40:36.6  -26:31:04.1& 24.9&  22.5&   4.4&   4.4\\
      238& 11:40:55.3  -26:30:43.5& 26.5&  97.5&   3.7&   3.4\\
      286& 11:40:47.9  -26:30:31.5& 27.5& 265.0&   3.1&   2.6\\
      297& 11:40:54.9  -26:30:29.8& 27.5& 497.6&   3.8&   3.4\\
      301& 11:40:45.2  -26:30:27.7& 27.4& 139.9&   4.0&   2.0\\
      361& 11:40:36.9  -26:30:08.8& 25.0&  23.6&   3.8&   4.3\\
      365& 11:40:58.1  -26:30:09.4& 25.0&  21.8&   3.5&   3.7\\
      366& 11:40:51.6  -26:30:08.0& 25.0&  29.5&   4.1&   5.0\\
      417& 11:40:45.3  -26:29:49.1& 24.7&  18.9&   3.8&   4.6\\
      441& 11:40:57.4  -26:29:38.3& 24.9&  83.9&   4.9&  13.5\\
      465& 11:40:48.0  -26:29:37.0& 26.8& 119.2&   3.5&   3.2\\
      470& 11:40:59.7  -26:29:35.0& 25.8&  49.9&   4.2&   3.9\\
      479& 11:40:51.0  -26:29:31.1& 25.3&  25.0&   3.3&   3.3\\
      484& 11:40:45.5  -26:29:30.0& 27.5&  96.6&   3.0&   1.4\\
  $^a$491& 11:40:48.2  -26:29:09.5& 21.6& 108.4&  68.3&
                                      \makebox[1ex][r]{341.9}\\
  $^a$515& 11:40:48.5  -26:29:07.3& 22.7&  48.9&  23.9&  64.3\\
      522& 11:40:49.4  -26:29:09.6& 23.7&  36.0&   8.6&  20.0\\
      561& 11:40:46.2  -26:29:03.2& 27.5&
                       \makebox[1ex][r]{9749.7}&   7.4&   6.8\\
      565& 11:40:56.6  -26:26:13.6& 25.3&  28.2&   3.0&   3.9\\
      674& 11:40:46.0  -26:28:35.7& 27.5& 269.4&   3.4&   2.6\\
      675& 11:40:55.3  -26:28:24.3& 26.8& 117.8&   3.9&   3.1\\
      703& 11:40:43.7  -26:28:21.8& 26.4&  86.7&   3.4&   3.4\\
      728& 11:40:36.9  -26:28:03.3& 23.7&  20.2&   7.1&  11.9\\
      739& 11:40:57.4  -26:27:07.8& 24.4&  20.6&   4.5&   6.3\\
  $^b$778& 11:41:02.4  -26:27:45.1& 24.1& 314.2&  27.4&  67.6\\
      877& 11:40:54.0  -26:28:01.1& 25.3&  31.6&   3.4&   4.3\\
      891& 11:40:59.1  -26:28:10.5& 25.0&  25.4&   4.4&   4.7\\
      941& 11:40:44.0  -26:28:33.5& 25.6&  32.0&   3.1&   3.3\\
      942& 11:40:49.8  -26:28:29.0& 26.2&  55.3&   3.3&   2.9\\
  $^c$968& 11:40:39.8  -26:28:45.4& 24.2&  26.6&   8.6&   9.8\\
  $^a$971& 11:40:47.8  -26:29:08.2& 25.5& 334.6&  10.6&  19.2\\
      980& 11:40:42.3  -26:28:51.3& 26.3& 113.1&   3.7&   4.7\\
     1009& 11:40:49.5  -26:29:11.1& 25.1&  57.1&   5.9&   8.5\\
     1017& 11:40:48.3  -26:29:01.7& 24.2&  68.2&   8.7&  21.5\\
\hline \hline
\end{tabular}
\end{center}
%\begin{center}
\footnotesize \noindent Notes: (1) Catalog number (2) Right ascension
and declination in J2000 coordinates (3) $B$ band magnitude (4) Rest
frame equivalent width (5) Significance of excess flux (6) Continuum
subtracted narrow band flux in $10^{-17}$ erg cm $^{-2}$
s$^{-1}$. Notes to individual objects: $^a$ Inside \lya\ halo of \rg,
$^b$ QSO described in Sect.~\ref{X-ray}, $^c$ QSO discovered in Paper
II.
%\end{center}
\end{table}

\begin{table}[htbp]
\caption{Extremely red objects ($I - K > 4.3$)}
\begin{center}
\begin{tabular}{r r r@{$\pm$}l r@{$\pm$}l} \hline \hline \label{erotable}
ID &  \multicolumn{1}{c}{Coordinates} & \multicolumn{2}{c}{$K$}
& \multicolumn{2}{c}{$I - K$} \\
(1) & \multicolumn{1}{c}{(2)} & \multicolumn{2}{c}{(3)} & 
\multicolumn{2}{c}{(4)} \\ \hline
       53& 11 40 44.1  -26 30 09.7& 21.3&   0.2&   4.4&   0.3\\
       72& 11 40 55.8  -26 30 05.8& 20.1&   0.1&   5.2&   0.3\\
       95& 11 40 44.2  -26 29 56.2& 22.0&   0.3&   4.8&   0.4\\
      111& 11 40 53.1  -26 29 50.3& 20.1&   0.1&   4.4&   0.1\\
      121& 11 40 56.5  -26 29 44.6& 19.2&   0.1&   4.3&   0.1\\
      151& 11 40 47.5  -26 29 41.3& 21.5&   0.2&   5.2&   0.6\\
      161& 11 40 50.4  -26 29 37.7& 20.9&   0.1&   4.8&   0.2\\
      163& 11 40 47.6  -26 29 38.2& 21.5&   0.2&   5.9&   1.0\\
      166& 11 40 50.4  -26 29 37.0& 20.7&   0.1&   5.2&   0.3\\
      171& 11 40 50.7  -26 29 33.8& 20.3&   0.1&   4.9&   0.2\\
      176& 11 40 50.7  -26 29 32.5& 20.2&   0.1&   4.7&   0.1\\
      189& 11 40 46.5  -26 29 27.1& 19.2&   0.0&   4.9&   0.1\\
      197& 11 41 00.1  -26 29 28.3& 20.7&   0.1&   7.3&   1.3\\
      210& 11 40 55.1  -26 29 24.7& 19.8&   0.1&   4.5&   0.1\\
      215& 11 40 51.1  -26 29 24.3& 20.9&   0.1&   5.6&   0.4\\
      226& 11 40 44.5  -26 29 20.8& 20.2&   0.1&   5.0&   0.1\\
  $^a$229& 11 40 48.4  -26 29 08.8& 16.1&   0.0&   4.7&   0.0\\
      233& 11 40 51.2  -26 29 21.4& 21.4&   0.3&   5.0&   0.6\\
      240& 11 40 52.1  -26 29 19.5& 20.9&   0.2&   4.7&   0.3\\
      248& 11 40 52.3  -26 29 16.9& 21.7&   0.3&   4.6&   0.5\\
      255& 11 40 48.7  -26 29 04.1& 21.7&   0.2&   5.8&   0.9\\
      256& 11 41 00.8  -26 29 04.7& 21.0&   0.2&   5.2&   0.4\\
      259& 11 40 51.1  -26 29 05.4& 21.1&   0.2&   4.4&   0.3\\
      267& 11 40 59.0  -26 28 57.6& 21.9&   0.3&   4.3&   0.4\\
      270& 11 40 44.3  -26 29 07.2& 19.2&   0.1&   5.3&   0.1\\
      272& 11 40 49.6  -26 29 07.7& 20.1&   0.1&   6.0&   0.4\\
      282& 11 40 47.8  -26 29 09.8& 20.4&   0.1&   4.6&   0.2\\
      284& 11 40 46.7  -26 29 10.4& 20.1&   0.1&   4.4&   0.1\\
      308& 11 41 01.2  -26 28 59.1& 22.0&   0.3&   4.7&   0.5\\
      343& 11 40 51.6  -26 28 51.7& 21.0&   0.2&   4.6&   0.3\\
      344& 11 40 49.8  -26 28 54.7& 19.9&   0.1&   5.8&   0.2\\
      348& 11 40 51.4  -26 28 53.3& 20.5&   0.1&   5.1&   0.3\\
      351& 11 40 51.7  -26 28 49.6& 21.4&   0.2&   4.8&   0.4\\
      359& 11 40 50.3  -26 28 49.8& 21.0&   0.1&   5.0&   0.3\\
      380& 11 40 43.5  -26 28 45.0& 21.1&   0.2&   5.1&   0.5\\
      402& 11 40 46.4  -26 28 35.8& 22.1&   0.3&   4.3&   0.4\\
      442& 11 40 44.1  -26 28 21.4& 21.2&   0.1&   5.1&   0.3\\
      460& 11 40 57.0  -26 28 17.4& 18.4&   0.1&   4.5&   0.1\\
      463& 11 40 45.5  -26 28 10.4& 21.4&   0.2&   5.1&   0.5\\
      472& 11 40 45.8  -26 28 12.8& 20.8&   0.1&   4.5&   0.2\\
      505& 11 40 54.8  -26 28 03.6& 20.7&   0.1&   4.4&   0.2\\
      506& 11 40 55.4  -26 28 00.4& 20.6&   0.1&   4.4&   0.2\\
      509& 11 40 47.4  -26 27 59.5& 20.2&   0.1&   4.8&   0.1\\
      531& 11 40 54.1  -26 27 48.1& 21.6&   0.2&   6.5&   1.1\\
\hline \hline
\end{tabular}
\end{center}
%\begin{center}
\footnotesize \noindent Notes: (1) Catalog number (2) Right ascension
and declination in J2000 coordinates (3) $K$ band magnitude and error
(4) $I - K$ colour and error. Notes to individual objects: $^a$ \rg.
%\end{center}
\end{table} % Note: manually replace 0.0 errors with 0.1

\end{document}